%% file: main.tex
\documentclass[runningheads]{llncs}

\usepackage{subfig}
\usepackage{graphicx}
\usepackage{amssymb,amsmath}
\usepackage{times,graphicx,url}
\usepackage{color,setspace,enumitem}
\usepackage{epstopdf}
\usepackage{caption}

\usepackage[ruled]{algorithm}
\usepackage{algpseudocode}
\usepackage{multicol}
\usepackage{soul}

\usepackage{hyperref}

\usepackage[binary-units=true]{siunitx}
\usepackage{tikz}

\input{nn_macros}
\input{macros}

\newcommand{\thc}[1]{{\color{blue}#1}}

\renewcommand{\af}[1]{{#1}}
\newcommand{\nn}[1]{{#1}}
\newcommand{\CG}[1]{{#1}}

\newcommand{\frfs}{{\sc CoBFS}}
\newcommand{\vmwABD}{{\sc CoABD}}
\begin{document}
\title{Fragmented Objects: Boosting Concurrency of Shared Large Objects\thanks{Supported by the Cyprus Research and Innovation Foundation under the grant agreement POST-DOC/0916/0090.}}

%
%

\author{
Antonio Fern\'andez Anta \inst{1}
\and
Chryssis Georgiou \inst{2}
\and
Theophanis Hadjistasi \inst{3}
\and
Nicolas Nicolaou \inst{3}
\and
Efstathios Stavrakis \inst{3}
\and
Andria Trigeorgi \inst{2}
}

\authorrunning{A. F. Anta et al.}
%

\institute{IMDEA Networks Institute, Madrid, Spain, 
\email{antonio.fernandez@imdea.org}\\
\url{} 
\and
University of Cyprus, Nicosia Cyprus\\
\email{\{chryssis, atrige01\}@cs.ucy.ac.cy}
\and
Algolysis Ltd, Limassol, Cyprus\\
\email{\{theo, nicolas, stathis\}@algolysis.com}}

\maketitle              
\begin{abstract}
This work examines strategies to handle \emph{large} shared data objects in distributed storage systems (DSS), while
boosting the number of concurrent accesses, 
maintaining strong consistency guarantees, and ensuring good operation performance. To this respect, we define 
the notion of \emph{fragmented objects:} concurrent objects composed of a list of fragments (or \emph{blocks}) that allow operations to manipulate each of their  fragments individually. 
As the fragments belong to the same object, it is not enough that each fragment is linearizable to 
have useful consistency guarantees in the composed object. Hence, we capture the consistency semantic 
of the whole object with the notion of \emph{fragmented linearizability}.
Then, considering that a variance of linearizability, \emph{coverability}, is more
suited for versioned objects like files, we provide an implementation of 
a distributed file system, called \frfs{}, that utilizes coverable fragmented objects (i.e., files). In \frfs{}, each file is a linked-list of coverable block objects. 
Preliminary emulation of \frfs{} demonstrates the potential of our approach in boosting the concurrency of strongly consistent large objects.

\end{abstract}

\input{introduction.tex}

\input{model_v1.tex}

\input{fragmented_objects_v2.tex}
\input{consistency_v2.tex}



\input{implementation_v2.tex}

\input{evaluation.tex}

\input{conclusions.tex}

%
%

\input{appendix_v1}

\end{document}

%% file: nn_macros.tex
\makeatletter
\def\mainlistofsymbols{
  \normalsize
  \vspace*{1.5 em}
  \@starttoc{los}
}

\def\partonelistofsymbols{
  \normalsize
  \vspace*{1.5 em}
  \@starttoc{p1los}
}

\def\parttwolistofsymbols{
  \normalsize
  \vspace*{1.5 em}
  \@starttoc{p2los}
}

\def\l@symbol#1#2{\addpenalty{-\@highpenalty} \vskip 4pt plus 2pt
{\@dottedtocline{0}{0em}{8em}{#1}{#2}}}
\makeatother

\newcommand{\newhiddensym}[2]{%
}

\newcommand{\stateSet}[1]{states(#1)}

\newcommand{\actionSet}[1]{actions(#1)}

\newcommand{\algIOA}[2]{\ifmmode{\text{#1}_{#2}}\else{$\text{#1}_{#2}$}\fi}

\newcommand{\EX}{\ifmmode{\xi}\else{$\xi$}\fi}
\newcommand{\EXF}{\ifmmode{\phi}\else{$\phi$}\fi}
\newcommand{\extend}[2]{#1\circ#2}

\newcommand{\acts}{\alpha}

\newcommand{\hist}[1]{H_{#1}}

\newcommand{\inter}[1]{
	\ifmmode{\left(\bigcap_{\mathcal{Q}\in#1}\mathcal{Q}\right)}
	\else{$\left(\bigcap_{\mathcal{Q}\in#1}\mathcal{Q}\right)$}
	\fi
}

\newcommand{\idSet}{\mathcal{I}}
\newcommand{\cSet}{\mathcal{C}}
\newcommand{\wSet}{\mathcal{W}}

\newcommand{\srvSet}{\mathcal{S}}

\newcommand{\verSet}{\mathit{Versions}}

\newcommand{\op}{\pi}

\newcommand{\rd}{\rho}

\mathchardef\mhyphen="2D
\newcommand{\trw}[2]{\act{tr-write}(#1)[#2]}

\newcommand{\pr}{p}

\newcommand{\bef}{\rightarrow}

\newcommand{\vid}[1]{\ifmmode{\nu_{#1}}\else{$\nu_{#1}$}\fi}

\newcommand{\seen}{\ifmmode{seen}\else{$seen$}\fi}

\newcommand{\fileSet}{\mathcal{F}}
\newcommand{\blockSet}{\mathcal{B}}
\newcommand{\valSet}{V}

\newcommand{\val}[1]{val_{#1}}

\newcommand{\maxts}[1]{\ifmmode{maxTS_{#1}}\else{$maxTS_{#1}$}\fi}
\newcommand{\maxtag}[1]{\ifmmode{maxTag_{#1}}\else{$maxTag_{#1}$}\fi}
\newcommand{\maxpair}[1]{\ifmmode{maxMPair_{#1}}\else{$maxMPair_{#1}$}\fi}
\newcommand{\mintag}[1]{\ifmmode{minTag_{#1}}\else{$minTag_{#1}$}\fi}
\newcommand{\maxps}{\ifmmode{maxPS}\else{$maxPS$}\fi}
\newcommand{\conftg}[1]{\ifmmode{confirmed_{#1}}\else{$confirmed_{#1}$}\fi}
\newcommand{\maxconftag}{\ifmmode{\ms{maxCT}}\else{$maxCT$}\fi}

%% file: macros.tex
\newcommand{\prf}[1]{{}}

\newtheorem{Def}{Definition}[section]

\newcommand{\sacode}[5]
{ \vspace{.06in} \hrule \vspace{.06in} 
 \noindent {\bf #1}: \\
 \footnotesize \noindent {\bf Signature:}\B \nobreak
 \normalsize \begin{quote} \nobreak #2 \end{quote}
 \footnotesize \noindent {\bf States:}\B \nobreak
 \begin{quote} \nobreak #3 \end{quote}
 \noindent {\bf Transitions:} \nobreak
 \vspace{-.2in} \nobreak
 \normalsize #4
 \vspace{-.06in} \hrule \vspace{.06in} 
}

\newcommand{\act}[1]{%
    \relax\ifmmode
        \mathord{\mathcode`\-="702D\sf #1\mathcode`\-="2200}%
    \else
        $\mathord{\mathcode`\-="702D\sf #1\mathcode`\-="2200}$%
    \fi
}

\newcommand{\tup}[1]{%
    \relax\ifmmode
      \langle #1 \rangle%
    \else
        $\langle$#1$\rangle$%
    \fi
}

\newcommand{\seq}[1]{%
    \relax\ifmmode
      \langle \! \langle #1 \rangle \! \rangle%
    \else
        $\langle \! \langle$ #1 $\rangle \! \rangle$%
    \fi
}

\newcommand{\B}{\vspace*{-\smallskipamount}}

\newcommand{\Nat}{{\N}}
\newcommand{\N}{\mathbb N}

\newcommand{\ms}[1]{%
    \relax\ifmmode
        \mathord{\mathcode`\-="702D\it #1\mathcode`\-="2200}%
    \else
        {\it #1}%
    \fi
}

\newcommand{\lit}[1]{%
    \relax\ifmmode
        \mathord{\mathcode`\-="702D\sf #1\mathcode`\-="2200}%
    \else
        {\it #1}%
    \fi
}

\newcommand{\XDK}[1]{}
\newcommand{\remove}[1]{} 
\newcommand{\uselater}[1]{} 

\newcommand{\af}[1]{\textcolor{red}{{#1}}}



%% file: introduction.tex
\section{Introduction}

In this paper we deal with the storage and use of shared readable and writable data in unreliable distributed
systems. 
Distributed systems 
are subject to perturbations, which may
include failures (e.g., crashes) of individual computers, or delays
in processing or communication.
In such settings, large (in size) objects are difficult to handle. Even more challenging is to provide linearizable consistency guarantees to such objects.

%
%
Researchers usually break large objects into smaller linearizable building blocks, with their composition yielding the complete consistent large object. For example, a linearizable shared R/W memory is composed of a set of linearizable shared R/W  objects \cite{ABD96}. By design, those building blocks are usually independent, in the sense that changing the value of one does not affect the operations performed on the others,
 and that operations on the composed objects are defined in terms of operations invoked on the (smallest possible) building blocks. 
Operations on individual linearizable registers do not violate the consistency of the larger composed linearizable memory space.



Some large objects, however, cannot be decomposed into independent building blocks. For example,
a file object can be divided into \af{\emph{fragments} or \emph{blocks},}
so that write operations (which are still issued on the whole file) modify individual fragments. However,
the composition of these fragments does not yield a linearizable file object: it is unclear how to order writes on the file when those are applied on different blocks concurrently.
At the same time, it is practically inefficient 
to handle large objects as single objects and use traditional algorithms 
(like the one in \cite{ABD96}) to distribute it consistently. 

\noindent\textbf{Related work:}
Attiya, Bar-Noy and Dolev \cite{ABD96},
proposed an algorithm, {colloquially
referred to as ABD}, that emulates a distributed shared R/W register 
in message-passing, crash-prone, 
asynchronous environments.
To ensure availability, the object is replicated among a set of 
servers and to provide operation ordering, a logical timestamp is associated with each written value. 
ABD tolerates replica server crashes, provided a majority of servers do not fail.
%
%
 Write operations
involve a single communication round-trip. The
writer broadcasts its request to all servers and it terminates once
it collects acknowledgments from some majority of servers.
A read involves two round-trips.
In the first, the reader broadcasts a request to all servers, collects acknowledgments
from some majority of servers, and it discovers the maximum timestamp. To ensure that any subsequent read will return a value associated with a timestamp at least as high as the discovered maximum, the reader propagates the value associated
with the maximum timestamp to at least a majority of servers before completion, forming the second round-trip.
 %
 ABD was later extended for the multi-writer/multi-reader model in \cite{LS97}, and its performance
 was later improved by several works, including \cite{CDGL04,GNS09,HNS17,FHN16,GHNS18}. Those solutions considered small objects, and relied on the dissemination of the object values in each operation, imposing a performance overhead when dealing with large objects. 

Fan and Lynch \cite{FL03} attempted to reduce performance overheads 
by separating the metadata of large objects from their value. In this way, communication-demanding operations were performed on the metadata, and large objects were transmitted to a limited number of hosts, and only when it was ``safe'' to do so. Although this work improved the latency of operations, compared to traditional approaches like \cite{ABD96,LS97}, it still required to transmit the entire large object over the network per read and write operation. Moreover, if two concurrent write operations affected different ``parts'' of the object, only one of them would prevail, despite updates not being directly ``conflicting.''

Recently, Erasure-Coded (EC) approaches have gained momentum and have proved being extremely 
effective in saving storage and 
communication costs, while maintaining strong consistency and fault-tolerance
\cite{CT06,CadambeLMM17,DGL08,SODA2016,radon,GIZA2017,Zhang2016,ARES19}.
EC approaches rely on the division of a shared object into coded blocks and deliver a single block to each data server. While 
very appealing for handling large objects, they face the challenge of
efficiently encoding/decoding data.
%
Despite being subdivided into several fragments, reads and writes 
are still applied on the entire object value. 
Therefore, multiple writers cannot work simultaneously on different parts of an object.

Value continuity 
is important when considering large objects, 
oftentimes overseen by distributed shared object implementations. 
In files, for example, a write operation should extend the latest written 
version of the object, and not overwrite any new value. 
\emph{Coverability} was introduced in \cite{NFG16} as a consistency 
guarantee that extends linearizability and concerns versioned objects. 
An implementation of 
a coverable (versioned) object was presented, where ABD-like reads 
return both the version and the value of the object. Writes, on the other hand, attempt to write a ``versioned'' value on the object. If the reported version is older than the latest,
then the write does not take effect and it is converted into a read operation, 
preventing
overwriting a newer version of the object.

\noindent\textbf{Contributions:} In this work we set the goal to study and formally define the 
consistency guarantees we can provide when fragmenting 
a large R/W object into smaller objects (blocks), so that operations are still issued on the former but are applied
on the latter. 
%
%
%
In particular, the contributions of this paper are as follows:

\begin{itemize}[leftmargin=5mm]
    \item We  define two types of concurrent objects: (i) the \emph{block} object, and (ii) the \emph{fragmented} object. Blocks are treated as  R/W objects, while fragmented objects are defined as lists of block objects (Section \ref{sec:fragment}).  
    
    \item 
    We examine the consistency properties 
    when allowing \af{R/W operations on} individual 
    blocks of the fragmented object, in order to enable concurrent modifications. 
    %
    %
    Assuming that each block is linearizable, we 
    define the precise consistency that the 
    fragmented object 
    provides, termed \emph{Fragmented Linearizability} (Section \ref{sec:atomicity}).
    
    
    \item We provide an algorithm that implements coverable fragmented objects.
    Then, we use it to build a prototype implementation of a distributed file system, called \frfs{}, by representing each file as a linked-list of coverable block objects. \frfs{} adopts a modular architecture, separating the object fragmentation process from the shared memory service,
    which allows
    to follow different fragmentation strategies and shared memory implementations. 
    We show that 
    \frfs{}
    preserves the validity of the fragmented object and satisfies \emph{fragmented coverability} (Section \ref{sec:fco}).
    
    \item 
    We describe an experimental development and deployment of \frfs{} on the 
    Emulab 
    testbed~\cite{emulab}. 
    Preliminary 
    results are presented, comparing our proposed algorithm to its non-fragmented counterpart. 
    Results 
    suggest
    that a fragmented object implementation 
    boosts concurrency while 
    reducing the latency of operations 
    (Section \ref{sec:evaluation}).
\end{itemize}


%% file: model_v1.tex
\section{Model}
\label{sec:model}
We are concerned with the implementations of highly-available 
replicated concurrent objects that support a set of operations. The
system is a collection of crash-prone, asynchronous processors with unique identifiers (ids)
from a totally-ordered set $\idSet$, composed of two main disjoint sets of processes: 
(a) a set $\cSet{}$ of client processes ids that may perform operations on a replicated object,
and (b) a set $\srvSet$ of server processes ids that each holds a replica of
the object. {Let $\idSet = \cSet\cup\srvSet$.} 


Processors communicate by exchanging messages via asynchronous point-to-point \emph{reliable}\footnote{Reliability is not necessary for the correctness of the 
algorithms we present. It is just used for simplicity of presentation.} channels; messages may be reordered. 
%
%
%
%
Any subset of client processes and up to a minority of servers (less than $|\srvSet|/2$), may crash at any time in an execution.

\noindent\textbf{Executions, histories and operations:} An \textit{execution} $\xi$
of a distributed algorithm $A$ is an alternating sequence of 
\textit{states} and \textit{actions} of $A$ reflecting the evolution in real time of the execution. A history $H_\xi$ is the subsequence
of the actions in $\xi$. We say that an operation 
$\pi{}$ is \textit{invoked} {(starts)} in an execution $\xi$ {when} the \textit{invocation action}
of $\pi$ appears in $H_\xi$, and $\pi$ responds to the environment (ends or completes) when the 
\textit{response action} appears in $H_\xi$. An operation is \textit{complete} in $\xi$ when both its
invocation and \textit{matching} response actions appear in $H_\xi$ in that order. A history $H_\xi$ is \textit{sequential} if it starts with an invocation action and each invocation is immediately followed by its matching response; otherwise, $H_\xi$ is \textit{concurrent}. Finally, $H_\xi$ is \textit{complete} if every invocation in $H_\xi$ has a matching response in $H_\xi$ (i.e., each operation in $\xi$ is complete). We say that an operation $\pi$
\textit{precedes in real time} an operation $\pi'$ (or $\pi'$ \textit{succeeds in real time} $\pi$) in an execution $\xi$, denoted by $\pi\rightarrow \pi'$, if the response of
$\pi $ appears before the invocation of $\pi'$ in $H_\xi$.
Two operations are {\em concurrent} if neither precedes the other.

\remove{
\noindent\textbf{Consistency}: We 
consider
\textit{linearizable} \cite{HW90} read/write (R/W) objects.
A complete history $H_\xi$ is linearizable if there exists a partial order on the
operations in $H_\xi$, such that, 
it respects the real-time order $\rightarrow$ of operations, and is 
consistent with the semantics of operations: \vspace{-.5em}
\begin{itemize}
    \item  A read operation returns a value no older than the 
value written by the latest preceding write operation, and any preceding read operation. \vspace{-.5em}
\end{itemize}
}

\noindent\textbf{Consistency}: We 
consider
\textit{linearizable} \cite{HW90} R/W objects.
A complete history $H_\xi$ is linearizable if there exists some total order on the
operations in $H_\xi$ s.t. 
it respects the real-time order $\rightarrow$ of operations, and is 
consistent with the semantics of operations.

\remove{
: \vspace{-.5em}
\begin{itemize}
    \write 
    \item  A read operation returns the 
value written by the latest preceding write operation. \vspace{-.5em}
\end{itemize}
}

Note that  we use read and write in an abstract way: ($i$) write represents any operation that changes the state of the object, and ($ii$) read any operation that returns that state.

%% file: fragmented_objects_v2.tex
\section{Fragmented Objects}
\label{sec:fragment}

A {\em fragmented object} is a concurrent object (e.g., 
can be accessed concurrently by multiple processes) that \nn{is composed of a finite list of {\em blocks}}.
Section~\ref{block-object} formally defines the notion of a \emph{block}, and Section~\ref{fragmented-object} gives the formal definition of a \emph{fragmented object}.
%
\remove{
\thc{[Line 154: I do not like this oppening.. Starting with `informally' seems sketchy. Maybe I am wrong. We need to re-write this line and beef-up a bit. Additionally, I do not understand \emph{concurrent object}. I do know \emph{concurrent operations}, \emph{concurrent object programming} but i never heard \emph{concurrent object}, To avoid any confusion, either we define it in the model with a very simple definition ``an object that allows concurent operations on it'' or we modify the above sentence to: ``a fragmented object is an object that allows concurrent operations on it, such that it contains....'' I believe it will be much cleaner.]}
}

\subsection{Block Object}
\label{block-object}
\nn{A \emph{block} $b$ is a concurrent R/W object with a unique identifier from a set $\blockSet$.}
A block 
has a value $val(b)\in\Sigma^*$, extracted from an alphabet $\Sigma$. 
For performance reasons it is convenient to bound the block length. Hence,
\nn{we denote by $\blockSet^\ell\subset\blockSet$, the set that contains bounded length blocks, 
s.t. $\forall b\in \blockSet^\ell$ the length of $|val(b)|\leq \ell$.}
We use $|b|$ to denote the length of the value of $b$ when convenient.
An \emph{empty block} is a block $b$ whose value is the empty string $\varepsilon$, i.e., $|b| = 0$.
%
Operation $\act{create}(b,D)$ is used to introduce a new block $b\in\blockSet^\ell$, 
initialized with value $D$,
such that $|D| \leq \ell$. Once created, 
block $b$ supports the following two operations: (i) $\act{read}()_b$ that returns the value of the object $b$, and (ii)  $\act{write}(D)_b$ that sets the value of the object $b$ to $D$, where $|D| \leq \ell$.

\nn{A block object is linearizable if is satisfies the linearizability properties} 
\cite{Lynch1996,HW90} 
with respect to its $\act{create}$ (which acts as a $\act{write}$), $\act{read}$, and $\act{write}$ operations. 
Once created, a block object is an atomic register~\cite{Lynch1996} whose value cannot exceed a predefined 
length~$\ell$.

\subsection{Fragmented Object}
\label{fragmented-object}

\nn{A \emph{fragmented object} $f$ is a concurrent R/W object with a unique identifier from a set~$\fileSet$.} \nn{Essentially, a fragmented object is a \textit{sequence} of blocks from 
$\blockSet$, with a value 
$val(f) = \tup{b_0,b_1,\ldots,b_n}$, where $b_i\in\blockSet, \text{for } i \in [0,n]$. 
Initially, each fragmented object contains an empty block, i.e., $val(f)=\tup{b_0}$ with $val(b_0)=\varepsilon$.} 
\nn{We say that $f$ is \textit{valid} and $f\in\fileSet^\ell$
if $\forall b_i\in val(f)$, $b_i \in \blockSet^\ell$. }
%
Otherwise, $f$ is \textit{invalid}. 
%
%
\nn{Being a R/W object, one would expect that a fragmented object $f\in\fileSet^\ell$, for any $\ell$, supports the following operations:} 
\vspace{-.5em}

\begin{itemize}[leftmargin=10mm]
    \item  \sloppy{$\act{read}()_f$ returns the list 
    $\tup{val(b_0), \ldots,val(b_n)}$, where  
    $val(f)=\tup{b_0,b_1,\ldots,b_n}$}
    
    \item \sloppy{$\act{write}(\tup{D_0,\ldots,D_n})_f$, $|D_i| \leq \ell, \forall i\in[0,n]$, sets the value of $f$ to $\tup{b_0,\ldots,b_n}$ 
    s.t. $val(b_i)=D_i, \forall i\in[0,n]$.}
   
\end{itemize}

\nn{Having the \act{write} operation to modify the values of all blocks in the list
may hinder in many cases the concurrency of the object.}
%
%
For instance, consider the 
following execution $\xi$. Let $val(f)=\tup{b_0, b_1}$, $val(b_0)=D_0$, $val(b_1)=D_1$, and assume that $\xi$ contains two concurrent writes by two different clients,
one attempting to modify block $b_0$, and the other attempting to modify block $b_1$: $\op_1 = \act{write}(\tup{D'_0, D_1})_f$ and $\op_2 = \act{write}(\tup{D_0, D'_1})_f$, followed by a $\act{read}()_f$. 
By linearizability,
the read will return either the list written in $\op_1$ or in $\op_2$ on $f$ (depending on how the operations are ordered by the linearizability property). 
However, \nn{as blocks are independent objects, it would be expected that both writes could take effect, 
with $\op_1$ updating the value of $b_0$ and $\op_2$ updating the value of $b_1$.}
To this respect, we \nn{redefine the \act{write}} to 
only update {\em one} of the blocks of a
fragmented object. \nn{Since the \act{update} does not manipulate
the value of the whole object, which would include also new blocks to be written, it should allow the update of a block $b$ with a value $|D| > \ell$.
This essentially leads to the generation of new blocks in the sequence.
More formally, the \act{update} operation is defined as follows:} 

\vspace*{-0.4em}
\nn{
\begin{itemize}[leftmargin=10mm]
    \item $\act{update}(b_i, D)_f$ updates the value of block $b_i\in f$ such that:
    	\begin{itemize}[leftmargin=5mm]
			\item if $|D| \leq \ell$: sets $val(b_i) = D$;
		    \item if $|D| > \ell$: partition $D=\{D_0, \ldots, D_k\}$ such that $|D_j|\leq\ell,  \forall j \in [0,k]$, set $val(b_i) = D_0$ and create blocks $b_i^j$, for $j\in [1,k]$ with $val(b_i^j) = D_j$, so that $f$ remains valid.
		\end{itemize}
\end{itemize}
}

\nn{With the update operation in place, fragmented objects resemble store-collect objects presented in~\cite{attiya2020storecollect}. 
However, fragmented objects aim to minimize the communication overhead by exchanging individual blocks (in a consistent manner) instead of exchanging the list (view) of block values in each operation.}
Since the update operation only affects a block in the list of blocks of a fragmented object, it potentially allows for a higher degree of concurrency.
\nn{It is still unclear what are the consistency guarantees we can provide when allowing concurrent updates on different blocks to take effect.}
Thus, we will consider that only operations $\act{read}$ and $\act{update}$ are issued in
fragmented objects. Note that the list of blocks of a fragmented object cannot be reduced. The contents of a block can be deleted by invoking an $\act{update}$ with an empty value.

Observe that as a fragmented object is composed of block objects, its operations are implemented by using $\act{read}$, $\act{write}$, and $\act{create}$ block operations. 
The $\act{read}()_f$ performs a sequence of $\act{read}$ block operations 
(starting from block $b_0$ and traversing the list of blocks) to obtain and return the value of the fragmented object. 
Regarding $\act{update}$ operations, if $|D| \leq \ell$, then the $\act{update}(b_i, D)_f$ operation performs a write operation on the block $b_i$ as
$\act{write}(D)_{b_i}$. However, if $|D| > \ell$, then $D$ is partitioned into substrings $D_0,\ldots,D_{k}$ each of length at most $\ell$. The update operation modifies the value of $b_i$ as $\act{write}(D_0)_{b_i}$. Then, $k$ new blocks $b_i^1,\ldots,b_i^{k}$ are created as $\act{create}(b_i^j, D_j), \forall j \in [1,k]$, and are inserted in $f$ between $b_i$ and $b_{i+1}$ (or appended at the end if $i=|f|$).
The sequential specification of a fragmented object is defined as follows:

\begin{definition}[Sequential Specification]
\label{def:sspec}
	The \emph{sequential specification} of a fragmented object $f\in\fileSet^\ell$ over the complete sequential history $H$ is defined as follows. 
	Initially $val(f)=\tup{b_0}$ with $val(b_0)=\varepsilon$.
	If at the invocation action of an operation $\op$ in $H$ has $val(f)=\tup{b_0, \ldots,b_n}$ and $\forall b_i\in f, val(b_i)=D_i$, and $|D_i|\leq\ell$. Then:
	\begin{itemize}[leftmargin=7mm]
		\item  if $\op$ is a $\act{read}()_f$, then $\op$ returns 
		$\tup{val(b_0), \ldots,val(b_n)}$.
		At the response action of $\op$, it still holds that $val(f)=\tup{b_0, \ldots,b_n}$ and $\forall b_i\in f, val(b_i)=D_i$.
		

		\item if $\op$ is an $\act{update}(b_i, D)_f$ operation, $b_i\in f$, 
		then at the response action of $\op$, $\forall j\neq i, val(b_j)=D_j$, and
		\begin{itemize}[leftmargin=5mm]
		    \item if $|D| \leq \ell$: $val(f)=\tup{ b_0,\ldots,b_n}$, $val(b_i) = D$;
		    \item if $|D| > \ell$: $val(f)=\tup{b_0, \ldots , b_i, b_i^1,\ldots,b_i^{k}, b_{i+1}, \ldots ,b_n}$, such that
		    $val(b_i) = D^0$ and $val(b_i^j) = D^j, \forall j \in [1,k]$, where $D=D^0| D^1|\cdots| D^k$
		     and $|D^j| \leq \ell,
		    \forall j \in [0,k]$.\footnote{The operator ``$|$" denotes concatenation. The exact way $D$ is partitioned is left to the implementation.}
		\end{itemize}
		
	\end{itemize}
\end{definition}


%% file: consistency_v2.tex
\section{Fragmented Linearizability}
\label{sec:atomicity}

A fragmented object is linearizable if it satisfies both the \emph{Liveness} (termination) and \emph{Linearizability} (atomicity) 
properties \cite{Lynch1996,HW90}. A fragmented object implemented by a single linearizable block 
is trivially linearizable as well. 
Here, we focus on fragmented objects that may contain a list of multiple 
linearizable blocks, and consider only $\act{read}$ and $\act{update}$ operations.
As defined, $\act{update}$ operations are applied on single blocks, which allows multiple $\act{update}$ operations 
to modify different blocks of the fragmented object concurrently. 
Termination holds since \act{read} and \act{update} operations on the fragmented object always complete.
It remains to examine the consistency properties.
%




\noindent\textbf{Linearizability:} Let $H_\xi$ be a sequential history of $\act{update}$ and $\act{read}$
invocations and responses on a fragmented object $f$.
Linearizability~\cite{Lynch1996,HW90} 
provides the illusion that the fragmented object is accessed sequentially respecting the real-time order, even when operations are invoked concurrently~\!\footnote{Our formal definition of linearizability is adapted from~\cite{AW94}.}: 

\begin{definition}[Linearizability]
	\label{def:atomic}
	A fragmented object $f$ is {\em linearizable} if, given any complete history 
	$H$, there exists a permutation $\sigma$ of all actions in $H$ such that: \vspace{-.5em}
	\begin{itemize}[leftmargin=10mm]
		\item $\sigma$ is a sequential history and follows the sequential specification of
		$f$, and 
		\item for operations $\pi_1, \pi_2$, if $\pi_1\bef \pi_2$ in $H$, then $\pi_1$ appears before $\pi_2$ in $\sigma$. 
	\end{itemize}
\end{definition}

\remove{
\begin{definition}[Linearizability]
	\label{def:atomic}
	A fragmented object $f$ is {\em linearizable} if, given any complete history 
	$H$, there exists a permutation $\sigma$ of all actions in $H$ such that: \vspace{-.5em}
	\begin{itemize}[leftmargin=10mm]
		\item $\sigma$ is a sequential history and follows the $B$-sequential specification of
		$f$ for the set $B=\blockSet$ of all blocks, and 
		\item for every pair of operations $\pi_1, \pi_2$, if $\pi_1\bef \pi_2$ in $H$, then $\pi_1$ appears before $\pi_2$ in $\sigma$. 
	\end{itemize}
\end{definition}
}

Observe, that in order to satisfy Definition \ref{def:atomic}, the operations must be totally ordered. Let us consider again the sample execution
$\xi$ from \nn{Section \ref{sec:fragment}}. Since we decided not to use write operations, the execution changes as follows. Initially, 
$val(f)=\tup{b_0, b_1}$, $val(b_0)=D_0$, $val(b_1)=D_1$,
and then $\xi$ contains two concurrent update operations by two different clients,
one attempting to modify the first block, and the other attempting to modify the second block: $\op_1 = \act{update}(b_0, D'_0)_f$ and $\op_2 = \act{update}(b_1, D'_1)_f$ ($|D'_0| \leq \ell$ and $|D'_1| \leq \ell$), followed by a $\act{read}()_f$ operation. In this case, since both update operations operate on different blocks,
independently of how $\op_1$ and $\op_2$ are ordered in the permutation $\sigma$, the $\act{read}()_f$ operation will return $\tup{D'_0,D'_1}$. Therefore,
the use of these $\act{update}$ operations has increased the concurrency in the fragmented object.
 
\begin{figure}[t]
    \centering
    \begin{tabular}{c|c}
        \includegraphics[width=0.48\linewidth, totalheight=3.0cm]{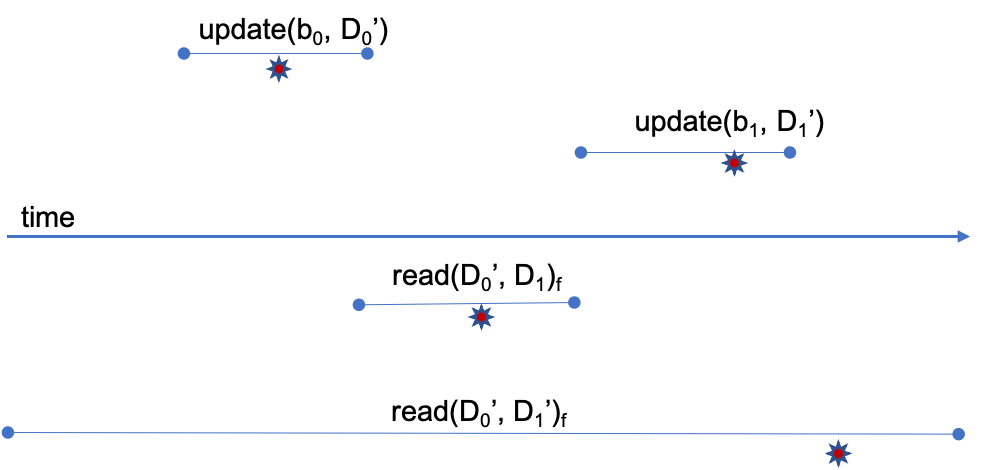} &
        \includegraphics[width=0.48\linewidth, totalheight=3.0cm]{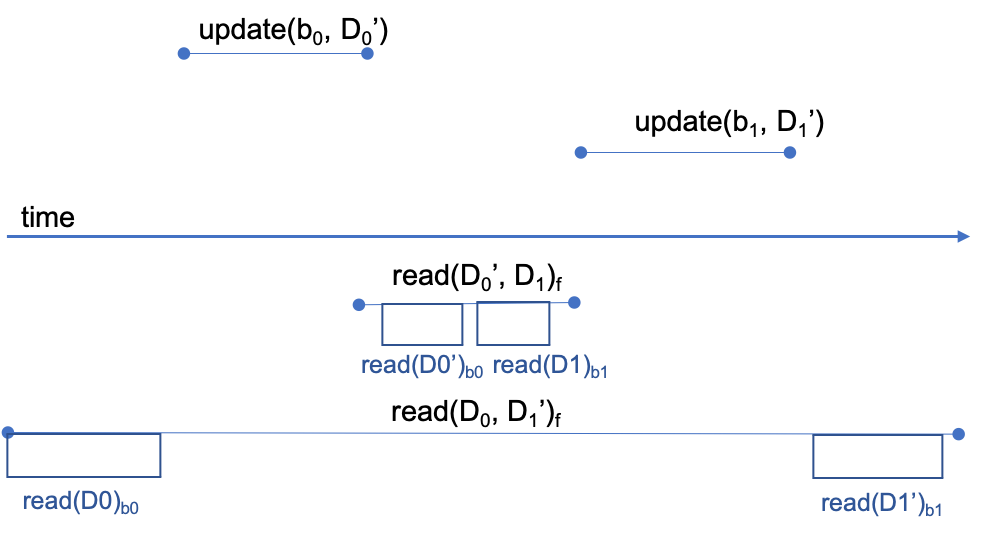}\\\\
        (a) 
         & 
        (b)
    \end{tabular}
    \caption{Executions showing the operations on a fragmented object. 
    Fig. (a) shows linearizable reads on the fragmented object (and serialization points), and (b) reads on the fragmented object that are implemented with individual linearizable reads on blocks.
    }
    \label{fig:linear}
\end{figure}

Using linearizable read operations on the entire fragmented object can ensure the linearizability of the fragmented object as can be seen in the example presented in Figure \ref{fig:linear}(a). However, providing a linearizable read when the object involves multiple R/W objects (i.e., an atomic snapshot) can be expensive or impact concurrency~\cite{Delporte-Gallet18}. Thus, it is cheaper to take advantage of the atomic nature of the individual blocks and invoke one read operation per block in the fragmented object. 
{\bf\em But, what is the consistency guarantee we can provide on the entire fragmented object in this case?}
As seen in the example of 
Fig.~\ref{fig:linear}(b), two reads concurrent with two update operations may violate linearizability on the {entire object}.
According to  the real time ordering of the operations on the individual blocks, block linearizability is preserved if the first read on the fragmented object should return $(D_0', D_1)$, while the second read returns $(D_0, D_1')$.
Note that we cannot find a  permutation on these concurrent operations that 
follows the sequential specification
of the {fragmented object}. 
%
%
Thus, the execution in Figure \ref{fig:linear}(b) violates linearizability.
\nn{This leads to the definition of \emph{fragmented linearizability} on the fragmented object, which relying on the fact that {\em each individual block is linearizable}, it allows executions like the one seen in Fig.~\ref{fig:linear}(b).} Essentially, fragmented linearizability captures the consistency one can obtain on a collection of linearizable objects, when these are accessed concurrently and individually, but under the ``umbrella" of the collection.

In this respect, we specify each $\act{read}()_f$ operation of a certain process, as a sequence of $\act{read}()_b$ operations on each block $b \in f$ by that process.  In particular, a read operation $\act{read}()_f$ that returns $\tup{val(b_0),\ldots,val(b_n)}$ is specified by $n+1$ individual read operations $\act{read}()_{b_0}$,..., $\act{read}()_{b_n}$,
that return $val(b_0)$, ...,
$val(b_n)$, respectively, where
$\act{read}()_{b_0} \bef ,\ldots, \bef \act{read}()_{b_n}$.
\remove{
Following this, the 
invocation action of 
$\act{read}()_f$ is the invocation action of 
$\act{read}()_{b_1}$ and the
response action of 
$\act{read}()_f$ is the response action of 
$\act{read}()_{b_n}$. Thus, from this point onward executions lead to histories consisting only of actions of operations on blocks.  
{\bf CG: Above we explain how to obtain histories with only operations on blocks. And then, below, we essentially reverse the process to obtain subhistories on the actions on operations on files. I think this is confusing. Is there a way we can unify the two?}
}

 Then, given a history $H$, we denote for an operation 
 $\op$ the history $\hist{}^\op$ which contains the
 actions extracted from $H$ and performed during $\op$ (including its invocation and response actions). Hence,
if $val(f)$ is the value returned by $\act{read}()_f$, then  $H^{\act{read}()_f}$ contains an 
invocation and matching response for a $\act{read}()_b$ operation, for each $b\in val(f)$. Then, from $H$, 
we can construct a history
$H|_f$ that only contains operations on the whole fragmented
object. In particular, $H|_f$ is the same as $H$ with the following changes: for each $\act{read}()_f$, if
$\tup{val(b_0),\ldots,val(b_n)}$ is the value returned by the
read operation, then we replace the invocation of 
$\act{read}()_{b_0}$ operation with the invocation of the 
$\act{read}()_f$ operation and the response of the 
$\act{read}()_{b_n}$ block with the response action for 
the $\act{read}()_f$ operation. Then we remove from $H|_f$
all the actions in $H^{\act{read}()_f}$.


\begin{definition}[Fragmented Linearizability]
\label{def:fragatomic}
Let $f\in\fileSet^\ell$ be a fragmented object, $H$ a complete history
on $f$, and $val(f)_H\subseteq\blockSet$ the value of $f$ at the 
end of $H$.
Then, $f$ is \emph{fragmented linearizable} if 
there exists a permutation $\sigma_b$ over all the actions on $b$ in $H$, 
$\forall b\in val(f)_H$, 
such that:\vspace{-.5em} 
	\begin{itemize}[leftmargin=10mm]
		\item $\sigma_b$ is a sequential history that 
		follows the sequential specification of $b$ \footnote{The sequential specification of a block is similar to that of a R/W register~\cite{Lynch1996}, whose value has bounded length.}, and 
		\item for operations $\pi_1, \pi_2$ that appear in $H|_f$ extracted from $H$, if $\pi_1\bef \pi_2$ in $H|_f$, then all operations on $b$ in  $H^{\op{1}}$ appear before any operations on $b$ in $H^{\op{2}}$ in $\sigma_b$. 
	\end{itemize}
\end{definition}

Fragmented linearizability guarantees that all concurrent operations on different blocks prevail, and only concurrent operations on the same blocks are conflicting.
%
Consider two 
reads
$r_1$ and $r_2$,
s.t.
$r_1 \bef r_2$;
%
then $r_2$ must return a supersequence of blocks with respect to the sequence returned 
by $r_1$, and that
 for each block belonging in both sequences, its value returned by $r_2$ is the same or newer than the one returned by $r_1$. 
%

\remove{
Consider the following simple execution: We have three concurrent writes on three blocks, $b_1, b_2,$ and $b_3$. Then two readers could return the following two sequences: $s_1=(b'_1, b_2$, and $b'_3)$ and
$s_2=(b_1, b'_2$, and $b_3)$. Notice that we cannot order $s_1$ and $s_2$ in a reasonable way. Therefore, we consider the consistency property where all write operations are {\em totally ordered} wrt individual blocks. That is, for the same block, the writes are totally ordered, whereas for different blocks we only require a partial order. 
Notice, that with this notion, the two sequences of the example above are both valid, and totally ordered wrt the individual blocks.\\ 
{\bf A picture depicting this execution would be helpful.}
}

%% file: implementation_v2.tex
\newcommand{\file}{f}
\newcommand{\block}{b}

\section{Implementing Files as Fragmented Coverable Objects}
\label{sec:fco}
Having laid out the theoretical framework of Fragmented Objects,
we now present a prototype implementation of 
a Distributed File System, we call \frfs{}.

\nn{When manipulating files it is expected that a value update
builds upon the current value of the object. In such cases a writer should 
be aware of the latest value of the object 
(i.e., by reading the object) before updating it. 
In order to maintain this property in our implementation we utilize \emph{coverable linearizable} blocks
as presented in~\cite{NFG16}. Coverability extends linearizability 
with the additional guarantee that object writes succeed when associating 
the written value with the “current” version of the object. 
In a different case, a write operation becomes a read operation and 
returns the latest version and the associated value of the object. Due to 
space limitations we refer the reader to~\cite{NFG16} 
for the exact coverability properties.}

\nn{By utilizing coverable blocks, our file system provides \emph{fragmented coverability} as a consistency guarantee. 
}
In our prototype implementation we 
consider each object to be a plain text file, however the underlying theoretical formulation allows for extending this implementation to support any kind of large objects.

\noindent\textbf{File as a coverable fragmented object:} Each 
file is modeled as a fragmented object with its blocks being coverable objects. 
The file is implemented as a {\bf\em linked-list of blocks} with the first block being a special block $b_{g}\in\blockSet$, which we call the {\bf \emph{genesis block}}, and then each block having a pointer $ptr$ to its next block, whereas the last block has a null pointer. Initially each file contains only the genesis block; the genesis block contains special purpose \af{(meta)} data.
%
%
The $val(b)$ of $b$ is set as a tuple, $val(b)=\tup{ptr, data}$.

\begin{figure}[t]
\centering
\subfloat{\includegraphics[width=0.9\linewidth]{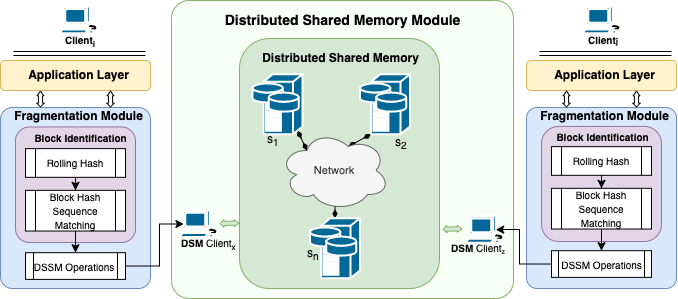}}
\caption{Basic architecture of \frfs{}}
\label{fig:architecture}
\end{figure}

\remove{
\begin{algorithm}[t]
			\caption{\small Signature and Operations of \frfs{}}
			\label{code:fs}
			\begin{multicols}{2}
			\begin{algorithmic}[1]
				\footnotesize
		       \Function{$\act{fs-read}$}{}$_{\file,\pr}$ 
				    \State return $\act{fm-read}()_{\file, \pr}$
				\EndFunction
			\Function{$\act{fs-write}$}{$\file$}$_{\pr}$ 
				    \State $\act{fm-block-identify}(\file)_{\pr}$
				\EndFunction
		\end{algorithmic}
		
	\end{multicols}
\end{algorithm}
}


\noindent\textbf{Overview of the Basic Architecure:} 
The basic architecture of \frfs{} appears in Fig.~\ref{fig:architecture}.
\frfs{} is composed of two main modules: ($i$) a Fragmentation Module (FM), and ($ii$) a Distributed Shared Memory Module (DSMM). In summary, the FM 
\nn{implements the fragmented object}
while the DSMM implements an interface to a shared memory service that allows read/write operations on 
individual block objects. Following this architecture, clients may access the file system through 
the FM, while the blocks of each file are maintained by servers through the DSMM. The FM uses the DSMM as an external service to write and read blocks to the shared memory. To this respect, \frfs{} is flexible enough to utilize any underlying distributed shared object algorithm.

\noindent\textbf{File and block id assignment:} A key aspect of our implementation is the 
unique assignment of ids to both fragmented objects (i.e. files) and individual blocks. 
A file $f\in\fileSet$ is assigned a pair $\tup{cfid,cfseq}\in\cSet\times\Nat$, where $cfid\in\cSet$ is the universally unique identifier of the client that created the file (i.e., the owner) and $cfseq\in\Nat$ is the client's local sequence number, incremented every time the client creates a new file and ensuring uniqueness of the objects created by the same client. 

In turn, a block $b\in \blockSet$ of a file is identified by a triplet $\tup{fid, cid, cseq}\in \fileSet\times\cSet\times\Nat$, where $fid\in\fileSet$ is the identifier of the file in which the block belongs to, $cid\in\cSet$ is the identifier of the client that created the block (this is not necessarily the owner/creator of the file), and $cseq\in\Nat$ is the client's local sequence number of blocks that is incremented every time this client creates a block for this file (this ensures the uniqueness of 
the blocks created by the same client for the same file).


\noindent\textbf{Distributed Shared Memory Module:} The DSMM implements a distributed R/W shared memory based on an \emph{optimized} {\em coverable variant} of the ABD algorithm, called \vmwABD{}~\cite{NFG16}.
\sloppy{The module exposes three operations for a block $b$: $\act{dsmm-read}_b$, $\act{dsmm-write}(v)_b$, and 
$\act{dsmm-create}(v)_b$.} The specification of each operation is shown in Algorithm \ref{code:smm}.
For each block $\block$, the DSMM maintains its latest known version $ver_\block$ and its 
associated value $val_\block$. Upon receipt of a read request for a block $\block$, the DSMM invokes a $\act{cvr-read}$ operation on $\block$
and returns the value received from that operation. 

\begin{algorithm}[t]
	\scriptsize
	\caption{\small DSM Module: Operations on a coverable block object $\block$ at client $\pr$}
	\label{code:smm}
	\vspace*{-5mm}
		\begin{multicols}{2}
		    
			\begin{algorithmic}[1]
			
				\State {\bf State Variables:}
				\State $ver_\block\in\Nat$ initially $0$; $val_\block\in\valSet$ initially $\bot$;  
			    \Statex
				\Function{$\act{dsmm-read}$}{ }$_{\block, \pr}$ 
				\State $\tup{val_\block, ver_\block} \gets \block.\act{cvr-read}()$ 
				\State {\bf return} $val_\block$
				\EndFunction

                \columnbreak
				\Function{$\act{dsmm-create}$}{$val$}$_{\block, \pr}$ 
				\State $\tup{val_\block, ver_\block} \gets \block.\act{cvr-write}(val, 0)$ 
				\EndFunction
                
				\Statex
				\Function{$\act{dsmm-write}$}{$val$}$_{\block, \pr}$
				\State $\tup{val_\block, ver_\block} \gets \block.\act{cvr-write}(val, ver_\block)$ 
				\State {\bf return} $val_\block$
				\EndFunction
		\end{algorithmic}
	\end{multicols}
	\vspace*{-4mm}
\end{algorithm}

\begin{algorithm}[t]
		\caption{\small Optimized coverable ABD (read operation)}
		\label{code:dsm}
		{\scriptsize
		\vspace*{-5mm}
		\begin{multicols}{2}
			\begin{algorithmic}[1]
				\State at each reader $r$ for object $b$
	        	\State {\bf State Variables:}
		        \State  $tg_b\in\N^+\times\wSet$ initially $\tup{0,\bot}$; 
		                $val_b\in V$, initially $\bot$
		        \Statex
		        \Function{\act{cvr-read}}{ }
        			\State {\bf send} $\tup{\text{{\sc read}},ver_b}$ to all servers\Comment{Query Phase} \label{line:reader:query}
        			\State {\bf wait until} $\frac{|\srvSet|+1}{2} $ servers reply 	
        			\State $maxP \gets \max(\{\tup{tg',v'} \text{ received from some server}\})$  \label{line:reader:maxtg}
        			\If {$maxP.tg > tg_b$}
        			    \State {\bf send} ($\text{{\sc write}}, maxP$) to all servers \Comment{Propagate Phase} \label{line:write:propagate}
        			    \State {\bf wait until} $\frac{|\srvSet|+1}{2} $ servers reply
        			    \State $\tup{tg_b, val_b} \gets maxP$
        			\EndIf
        			\State return($\tup{tg_b, val_b}$) 
        		\EndFunction
	            \Statex

            	\State at each server $s$ for object $b$
            	\State{\bf State Variables:}
            	\State  $tg_b\in\N^+\times\wSet$ initially $\tup{0,\bot}$; 
		                $val_b\in V$, initially $\bot$
		        \Statex
            	\Function{\act{rcv}}{$M$}$_{q}$\Comment{Reception of a message from $q$}
                    \If {$M.type\neq \text{{\sc read}}$ and $M.tg> tg_b$} 	\label{line:server:ts-comparison}
            			\State  $\tup{tg_b,val_b}\gets \tup{M.tg,M.v}$ \label{line:server:update}
            		\EndIf
            		\If {$M.type = \text{{\sc read}}$ and $M.tg \geq tg_b$} 
            		    \State  send($\tup{tg_b,\bot}$) to $q$ 	\label{line:server:reply:nocontent}
            		    \Comment{Reply without content}
            		\Else
            		    \State  send($\tup{tg_b,val_b}$) to $q$ 	\label{line:server:reply}
            		    \Comment{Reply with content}
            		\EndIf
            	\EndFunction
		\end{algorithmic}
	\end{multicols}
	\vspace*{-4mm}
	}
\end{algorithm}

To reduce the number of blocks transmitted per read, we apply a simple yet very effective optimization (Algorithm~\ref{code:dsm}): a read sends a {\sc read} request to all the servers including its local version in the request message. When a server receives a {\sc read} request it replies with both its local tag and block content only if the tag enclosed in the {\sc read} request is smaller than its the local tag; otherwise it replies with its local tag without the block content. 
Once the reader receives replies from a majority of servers, it detects the maximum tag among the replies, and checks if it is higher than the local known tag. If it is, then it forwards 
the tag and its associated block content to a majority of servers; if not then the read operation returns the locally known tag and block content without performing the second phase. 
While this optimisation makes a little difference on the non-fragmented version of the ABD (under read/write contention), 
it makes a significant difference in the case of the fragmented objects. 
For example, if each read is concurrent with a write causing the execution of a second phase, then the read sends the complete file to the servers; in the case of fragmented objects only the fragments that changed by the write will be sent over to the servers, resulting in significant reductions. 

The $\act{create}$ and $\act{write}$ operations
invoke $\act{cvr-write}$ operations to update the value of the shared block~$\block$. Their main 
difference is that version $0$ is used during a $\act{create}$ operation to indicate that
this is the first time that the block is written.
Notice that
the write in $\act{create}$ will always succeed as it will introduce a 
new, never before written block, whereas operation $\act{write}$ may be converted to a read operation,
thus retrieving and returning the latest value of $\block$.
We refer the reader to \cite{NFG16} for the implementation of $\act{cvr-read}$ and $\act{cvr-write}$, which 
are simple variants of the corresponding implementations of ABD~\cite{ABD96}. We state the following lemma:

\begin{lemma}
\label{lem:smm}
The DSMM implements R/W coverable block objects. 
\end{lemma}
\begin{proof}
    When both the read and write operations perform two phases the correctness 
of the algorithm is derived from Theorem~10 in~\cite{NFG16}. It is easy to 
    see that the optimization does not violate linearizability. The second 
    phase of a read is omitted when all the servers 
    reply with a tag smaller or equal to the local tag of the reader $r$. Since however,
    a read propagates its local tag to a majority of servers at every tag update, then every subsequent operation will observe (and return) the latest value of the object to be associated with a tag at least as high as the local tag of $r$.
\end{proof}

\vspace*{-0.4em}
\noindent\textbf{Fragmentation Module:}
The FM is the core concept of our implementation. Each client has a FM responsible for ($i$) fragmenting the file into blocks and identify modified blocks, and ($ii$) follow a specific strategy to store and retrieve the file 
blocks from the R/W shared memory. As we show later, the block update strategy followed by FM is
necessary in order to preserve the structure of the \nn{fragmented object} and sufficient to preserve the properties of fragmented coverability. For the file division of the blocks and the identification of the newly created blocks, the FM contains a \emph{Block Identification (BI) module} that utilizes known approaches for
data fragmentation and diff extraction.


\noindent\textbf{Block Identification (BI):} Given the data $D$ of a file $f$ the goal of BI is to
break $D$ into data blocks $\tup{D_0,\ldots,D_n}$, s.t. the size of each $D_i$ is less than a 
predefined upper bound $\ell$. Furthermore, by drawing ideas from 
the RSYNC (Remote Sync) algorithm~\cite{rsync},
given two versions of the same file, say $f$ and $f'$,
the BI tries to identify blocks that
($a$) may exist in $f$ but not in $f'$ \nn{(and vice-versa)}, 
or 
($b$) they have been changed from $f$ to $f'$. 
%
To achieve these goals BI proceeds in two steps: (1) it fragments $D$ into blocks, using the 
\emph{rabin fingerprints} rolling hash algorithm~\cite{rabin}, and (2) 
\nn{it compares the hashes of the blocks of the current 
and the previous version of the file} using a string matching
algorithm~\cite{stringMatching} to determine the modified/new data blocks.
The role of BI within the architecture of \frfs{} and its process flow appears in Fig.~\ref{fig:example}, while its specification is provided in Algorithm~\ref{code:BI}.
%
A high-level description of \emph{BI} has as follows:\vspace{-.5em}

\begin{figure}[t]
\centering
\subfloat{\includegraphics[width=0.9\linewidth]{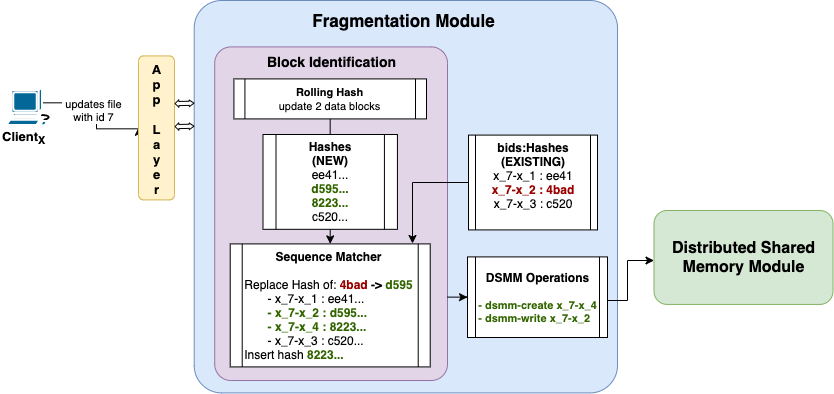}}
\caption{Example of a writer $x$ writing text at the beginning of the second block of a text file with id $f_{id}=7$. The hash value of the existing second block ``$4bad..$'' is replaced with ``$d595..$'' and a new block with hash value ``$8223..$'' is inserted immediately after. 
The block $b_{id} = $ x{\_}7-x{\_}2
and the new block $b_{id} = $ x{\_}7-x{\_}4 are sent to the DSM.}
\label{fig:example}
\vspace*{-5mm}
\end{figure}

\begin{algorithm}[t]
			\caption{\small Fragmentation Module: BI and Operations on a file $\file$ at client $\pr$}
			\label{code:BI}
			\vspace*{-5mm}
			\begin{multicols}{2}
			\begin{algorithmic}[1]
			\scriptsize{

			\State {\bf State Variables:}
			\State $H$ initially $\emptyset$; $\ell\in\Nat$;
			\State $\mathcal{L}_\file$ a linked-list of blocks,  initially $\tup{b_g}$;
			\State 
			$bc_\file\in\Nat$ initially $0$;  
		    \Statex
			\Function{$\act{fm-block-identify}$}{ }$_{\file, \pr}$ 
			    \State $\tup{newD,newH} \gets \act{RabinFingerprints}(\file,{\ell})$
			    \State $curH = hash(\mathcal{L}_\file)$
			    \State\Comment{hashes of the data of the blocks in $\mathcal{L}_\file$}
			    \State $C \gets \act{SMatching}(curH, newH)$
			
			    \State \Comment{modified} \label{fig:line:modify}
			    \For{$\tup{h(b_j), h_k} \in C.mods$ s.t. $h(b_j)\in curH, h_k\in newH$} 
			    \State $D\gets\{D_k: D_k\in newD \wedge h_k = hash(D_k)\}$
			    \State $\act{fm-update}(b_j, D)_{\file,\pr}$ \label{fig:line:modify:end}
			    \EndFor
			    
			    \State \Comment{inserted} \label{fig:line:insert}
			    \For{$S \in C.inserts$ s.t. $h_i\in S$ are in sequence}
			    \State $D\gets\{D_i: h_i\in S \wedge D_i\in newD \wedge h_i=hash(D_i)\}$
			    \State $b \gets b_j$ s.t. $\forall h_i\in S$ inserted after $h(b_j)$
			    \State $\act{fm-update}(b, D)_{\file,\pr}$ \label{fig:line:insert:end}
			    \EndFor
			        
			\EndFunction
			\Statex
		
		\columnbreak
			\Function{$\act{fm-read}$}{ }$_{\file, \pr}$ 
			    \State $\block \gets val(\block_{\CG{g}}).ptr$
			    \State $\mathcal{L}_f \gets \tup{b_g}$ \Comment{reset $\mathcal{L}_f$}
			    \While{$\block~not~$NULL}
			        \State $val(\block) \gets \act{dsmm-read}()_{\block,\pr}$
			         \State $\mathcal{L}_f.insert(val(\block))$
			        \State $\block \gets val(\block).ptr$
			    \EndWhile
				\State {\bf return} $\act{Assemble}(\mathcal{L}_f)$
			\EndFunction
			\Statex
				
			
			\Function{$\act{fm-update}$}{$\block, D=\tup{D_0, D_1,\ldots, D_k}$}$_{\file, \pr}$
			\For{$j=k : 1$}
			    \State $\block_j \gets \tup{\file,\pr,bc_\file\text{++}}$ \Comment{set block id}
    	        \State $val(\block_j).data = D_j$  \Comment{set block data}
    	        \If{$j<k$}
    	            \State $val(\block_j).ptr = \block_{j+1}$   \Comment{set block ptr}
    	        \Else
    	            \State $val(\block_j).ptr = val(\block).ptr$ 
    	            \State \Comment{point last to $\block$ ptr}
    	        \EndIf
    	        \State $\mathcal{L}_f.insert(val(b_j))$
    	        \State $\act{dsmm-create}(val(b_j))_{b_j}$
    	    \EndFor
    	    \State $val(\block).data = D_0$
    	    \If{$k > 0$}
    	        \State $val(\block).ptr = \block_{1}$ \Comment{change $\block$ ptr if $|D|>1$}
    	    \EndIf
    	    \State $\act{dsmm-write}(val(\block))_{\block}$
		\EndFunction
			
			}
		\end{algorithmic}
	\end{multicols}
	\vspace*{-3mm}
\end{algorithm}

\begin{itemize}[leftmargin=5mm]
  \item {\bf Block Division:} Initially, the BI partitions a given file $f$ into data blocks based on its contents, using \emph{rabin fingerprints}. This algorithm
  identifies the block boundaries and it performs content-based chunking by
 calculating and returning the fingerprints (block hashes) over a sliding window, 
and guarantees that each block identified has a bounded size of no more than $\ell$. 
  
  \item {\bf Block Matching:} Given the set of blocks $\tup{D_0,\ldots,D_m}$ and associated block hashes $\tup{h_0,\ldots,h_m}$ generated by the rabin fingerprint 
  algorithm, the BI tries to match each hash to a block identifier, based on the block ids produced during
  the previous division of file $\file$, say $\tup{b_0,\ldots,b_n}$. 
  \nn{We produce the vector $\tup{h(b_0),\ldots,h(b_n)}$ where $h(b_i) = hash(val(b_i).data)$ from the current blocks of $f$, and using 
  a string matching algorithm~\cite{stringMatching} we compare the two hash vectors to
  obtain one of the following statuses for each entry: ($i$) equal, ($ii$) modified, ($iii$) inserted, ($iv$) deleted.}

  \item {\bf Block Updates:} 
  Based on the hash statuses computed through block matching previously, the blocks of the fragmented object are updated.
  In particular, in the case of equality, if a $h_i = h(b_j)$ then  $D_i$ is identified as the data of block $b_j$. In case of modification, e.g. $(h(b_j), \, h_i$), an $\act{update}(b_j, \{D_i\})_{\file,\pr}$ action is then issued to modify the 
data of $b_j$ to $D_i$ (Lines \ref{fig:line:modify}:\ref{fig:line:modify:end}). \nn{In case new hashes (e.g. $\tup{h_i, h_k}$) 
are inserted after the hash of block $b_j$ (i.e. $h(b_j)$), then 
the action $\act{update}(b_j, \{val(b_j).data, D_i, D_k\})_{\file,\pr}$ is performed to create the new blocks 
after $b_j$ (Lines \ref{fig:line:insert}: \ref{fig:line:insert:end}).}
\nn{In our formulation block deletion is treated as a modification that sets an empty data value 
thus, in our implementation {\em no blocks are deleted}.}
\end{itemize}\vspace{-.5em}


\noindent\textbf{FM Operations:}
The FM's external signature includes the two main operations of a fragmented object: $\act{read}_\file$, and $\act{update}_\file$. Their specifications 
appear in Algorithm~\ref{code:BI}.

\noindent\textbf{\underline{Read operation - $\act{read}()_{\file,\pr}$:}}
To retrieve the value of a file $\file$, a client $\pr$ may invoke a $\act{read}_{\file,\pr}$ to the fragmented object. Upon receiving, the FM issues a series of reads on file's blocks; starting from the genesis block of $\file$ and proceeding to the last block by following the pointers in the linked-list of blocks comprising the file. All the blocks are assembled into one file via the \act{Assemble()} function. 
The reader $\pr$ issues a read for all the blocks in the file. This is done to ensure the
property stated in the following lemma:

\begin{lemma}
\label{lem:reads}
    Let $\EX$ be an execution of \frfs{} with  two reads $\rd_1 = \act{read}_{\file,p}$ and 
    $\rd_2 = \act{read}_{\file,q}$ from clients $p$ and $q$ on 
    the fragmented object $\file$, s.t. $\rd_1\bef\rd_2$. If $\rd_1$ returns a list
    of blocks $\mathcal{L}_1$ and $\rd_2$ a list $\mathcal{L}_2$, then $\forall \block_i\in\mathcal{L}_1$, then $\block_i\in\mathcal{L}_2$ and 
    $version(\block_i)_{\mathcal{L}_1} \leq version(\block_i)_{\mathcal{L}_2}$. 
\end{lemma}


\noindent\textbf{\underline{Update operation - $\act{update}(b, D)_{\file,\pr}$:}}
Here we expect that the update operation accepts a block id and a set of data blocks (instead of a single data object), since the division is performed by the BI module. Thus, $D=\tup{D_0,\ldots, D_k}$, 
for $k \geq 0$, with the size $|D|=\sum_{i=0}^k |D_i|$ and the size of each $|D_i|\leq\ell$ for some 
maximum block size $\ell$.
Client $\pr$ attempts to update the value of a block with identifier $\block$ in file $\file$ with the data in $D$. Depending on the size of $D$ the update operation will either perform a write on the block if $k=0$, or it will create new blocks and update the block pointers in case $k>0$. Assuming that $val(b).ptr=b'$ then:\vspace{-.5em}
%
%
\begin{itemize}[leftmargin=10mm]
\item \sloppy{$k=0$: In this case $\act{update}$, for block $b$, calls  $\act{write}(\tup{val(b).ptr,D_0},\tup{p,bseq})_b$.}  
\item $k>0$: 
Given the sequence of chunks $D=\tup{D_0,\ldots, D_k}$ the following block operations are performed in this particular order:
\begin{itemize}[leftmargin=6mm, label=$\rightarrow$]
	\item $\act{create}(b_{k}=\tup{f,p,bc_p\text{++}},\tup{b',D_{k}},\tup{p,0})$ 
	~ ~ ~ ~ {\footnotesize{** Block $b_{k}$ ptr points to $b'$ **}}
	
	\item $\ldots$
	
	\item $\act{create}(b_1=\tup{f,p,bc_p\text{++}},\tup{b_2,D_1},\tup{p,0})$ 
	~ ~ ~ ~ {\footnotesize{** Block $b_1$ ptr points to $b_2$ **}}
	
	\item $\act{write}(\tup{b_1,D_0},\tup{p,bseq})_b$
	~ ~ ~ ~ ~ ~ ~ ~ ~ ~ ~ ~ ~ ~ ~ ~ ~ ~ {\footnotesize{** Block $b$ ptr points to $b_1$ **}}
\end{itemize}
\end{itemize}

\nn{The challenge here was to insert the list of blocks without causing any concurrent operation to return a divided fragmented object, while also avoiding blocking any ongoing operations.} To achieve that, \act{create} operations are executed in a reverse order: we first create block $b_{k}$ pointing to $b'$, and we move backwards until creating $b_1$ pointing to block $b_2$. The last operation, \act{write}, tries to update the value of block $b_0$ with value $\tup{b_1,D_0}$. 
If the last coverable write completes successfully, then all the blocks are inserted in $f$ and the update is \emph{successful}; otherwise none of the blocks appears in $f$ and thus the update is \emph{unsuccessful}. 
This is captured by the following lemma:

\begin{lemma}
\label{lem:last:write}
In any execution $\EX$ of \frfs{}, if $\EX$ contains an $\op=\act{update}(b,D)_{\file,\pr}$,
then $\op$ is successful iff the operation $b.\act{cvr-write}$ called within  $\act{dsmm-write}(val(b))_{b,\pr}$, 
is successful.
\end{lemma}

\begin{proof}
It is easy to see that if $\op=\act{update}(b,D)_{\file,\pr}$ is successful, then all the 
$\act{dsmm-write}$ operations invoked within $\op$, including $\act{dsmm-write}(val(b))_{b,\pr}$, are successful. It remains to show that $\op$ can only by unsuccessful whenever $\act{dsmm-write}(val(b))_{b,\pr}$ is unsuccessful. In the case where $D$ contains a single chunk, i.e. $D=\tup{D_0}$ then $\op$ 
invokes a single $\act{dsmm-write}(val(b))_{b,\pr}$ with $val(b).data = D_0$. If the $\act{cvr-write}$
invoked in that operation is unsuccessful then $\op$ is also unsuccessful. In the case where $k>0$, 
$\op$ invokes $k-1$ create operations with new block identifiers (due to the incremented block counter $bc$). The $\act{cvr-write}$ operation on every such block will be successful as (i) the block id 
$\tup{f,\pr,bc}$ (and thus the block) can only be generated by process $\pr$, and (ii) the block 
is not yet inserted in the link-list. So no other write operation will attempt to $\act{cvr-write}$
the same block concurrently. So the only operation that may fail in this case as well, is the 
$\act{dsmm-write}(val(b))_{b,\pr}$ as $b$ was a part of the list and may be accessed concurrently
by a writer $q\neq \pr$.
\end{proof}

Now a read operation may return a list that contains a block $b_i$ only if $b_i$ was written 
by a successful update operation. More formally:

\begin{lemma}
\label{lem:update:block}
    In any execution $\EX$ of \frfs{}, if a $\rd = \act{read}_{\file,\pr}$ operation returns a 
    list $\mathcal{L}$ then for any block $b\in\mathcal{L}$ there exists successful $\act{update}(*)_{\file,*}$ operation that either precedes or is concurrent to $\rd$ and invokes $\act{sm-create}(val(b))_b$ operation.
\end{lemma}

\begin{proof}
According to our protocol it is clear that a block with id $b$ appears in the list of $f$ only if 
that is created and written during an $\act{update}_{f,*}$ operation. Also, if the block is created
by an $\act{update}$ that precedes $\rd$, then no other block in the list will point to $b$, $\rd$ will not invoke a $\act{sm-read}_b$ operation for $b$, and thus $b\notin\mathcal{L}$.

So it remains to examine the case where $\rd$ may obtain $b$ from an unsuccessful $\act{update}_{f,*}$. Let us assume by contradiction that a read operation may return a block $b$ for a file $f$ created by an unsuccessful update. Let $b\in\tup{b_1,\ldots, b_n}$, the list of blocks that the update needs to write on the DSM. In particular, the operation will create all the blocks $\tup{b_2,\ldots,b_n}$ and attempt to write block $b_1$. There are two cases to consider: ($i$) either $b$ is equal to $b_1$, or ($ii$) $b$ is in $\tup{b_2,\ldots, b_n}$. 

If case ($i$) is true, then $\pr$ 
will invoke a $\act{sm-write}(val(b))_b$ as $b$ is the block that is updated. However, since we 
assume that the update was not successful, then by Lemma \ref{lem:last:write}, the write 
operation is not successful. Thus, according to the coverable DSM, $b$ was never written and this
contradicts the assumption that $\pr$ obtain $b\in\mathcal{L}$.

If case ($ii$) holds, then $b$ was created by $\pr$ (an operation that cannot fail). However, since the update is not successful, then $b_1$ was not written in the list. It is also true that there is no
link path leading to $b$ since the only path was $b_1\bef b_2\bef\ldots\bef b$. So, during the traversal 
of the blocks, the read operation will not see $b_1$ and thus will never reach and obtain $b$, contradicting again our initial assumption.
\end{proof}

The above lemma will help us to show that the linked-list used for implementing our 
fragmented object stays connected in any execution. 

\begin{lemma}
\label{lem:list}
    In any execution $\EX$ of \frfs{}, if a $\act{read}_{\file,\pr}$ operation returns a 
    list $\mathcal{L}=\tup{b_g, b_1,\ldots,b_n}$ for a file $\file$, 
    then $val(b_g).prt = b_1$, $val(b_i).ptr = b_{i+1}$, for $1\leq i<n-1$, and $val(b_n).ptr = \bot$.
\end{lemma}

\begin{proof}
 Assume by contradiction that there exist some $b_i\in\mathcal{L}$, s.t. $val(b_i).ptr\neq b_{i+1}$ \CG{(or $val(b_g).prt \neq b_1$)}. By Lemma \ref{lem:update:block}, a block $b_i$ may appear in the list returned by a read operation only if it was created by a successful update operation,
 say w.l.o.g. $\op=\act{update}(b, D)_{\file,*}$. Let $D = \tup{D_0, \ldots, D_k}$ and $\blockSet = \tup{b_1,\ldots, b_k}$ 
 be the set of $k-1$ blocks created in $\op$, with $b_i\in \blockSet$.
 By the design of the algorithm we create a single linked path from $b$ to $b_k$, by pointing $b$ to 
 $b_1$ and each $b_j$ to $b_{j+1}$, for $1\leq j<k$. Block $b_k$ points 
 to the block pointed by $b$ at the invocation of $\op$, say $b'$. 
 So there exists a path $b\bef b_1\bef\ldots \bef b_i$ 
 that also leads to $b_i$. According again to the algorithm, $b_{j+1}\in \blockSet$ is created and written before $b_j$, for $q\leq j <k$. So when the $b_j.\act{cvr-write}$ is invoked, the operation 
 $b_{j+1}.\act{cvr-write}$ has completed, and thus when $b$ is written successfully all the blocks in 
 the path are inserted successfully in $f$. So, if now $b_i$ is different than $b_k$ by the construction of the update then both $b_i$ and $b_{i+1}$ are in the list with $val(b_i).ptr=b_{i+1}$ contradicting our assumption. 
 
 If now $b_i=b_k$, then $val(b_i).ptr = b'$. Since $b$ was pointing to $b'$ at the invocation 
 of $\op$ then $b'$ was either ($i$) created during the update operation that also created $b$, or ($ii$) was
 created before $b$. In case ($i$), by Lemma \ref{lem:last:write}, the update operation that created
 $b$ was successful and thus $b'$ must be created and inserted in $f$ as well. In case ($ii$) it follows that $b$ is the last inserted block of an update and is assigned to point to $b'$. With a simple induction one may show that the update operation that created $b'$ must precede the update that created $b$. 
 Since no block is 
 deleted, then $b'$ remains in $\mathcal{L}$ when $b_i$ is created and thus $b_i$ points to an existing block.
 Furthermore, since $\op$ was successful, then it successfully written $b$ and hence only the blocks 
 in $\blockSet$ were inserted between $b$ and $b'$ at the response of $\op$. So $b'$ must be the 
 next block after $b_i$ in $\mathcal{L}$ at the response of $\op$ and there is a path between $b$ and $b'$. This completes our proof.
\end{proof}

This leads us to the following: 

\begin{theorem}
     \frfs{} implements a R/W Fragmented Coverable object.
\end{theorem}     

\begin{proof}
 By Lemma \ref{lem:smm} every block operation in \frfs{} satisfies coverability and together with Lemma \ref{lem:reads} it follows 
 that \frfs{} implements
 a coverable fragmented object satisfying the properties presented in Definition \ref{def:fragatomic} 
Also, the BI ensures that the size of each block is limited under a bound $\ell$ 
 and Lemma \ref{lem:list} ensures that each operation obtains a connected list of blocks. Thus, \frfs{}
 implements a \emph{valid} fragmented object.~\end{proof}
 \vspace{-.5em}

%% file: evaluation.tex
\section{Preliminary Evaluation}
\label{sec:evaluation}



To further appreciate 
the proposed approach from an applied point of view, we performed a preliminary evaluation of \frfs{} against \vmwABD{}. Due to the design 
of the two algorithms, \vmwABD{} will transmit the entire file per read/update operation, while \frfs{} will transmit as many blocks as necessary 
for an update operation, but perform as many reads as the number of blocks during a read operation. The two algorithms use the read optimization 
of Algorithm~\ref{code:dsm}. 
Both were implemented and deployed on \emph{Emulab,}~\cite{White+:osdi02}, a network testbed with tunable and controlled environmental parameters.

%



\noindent\textbf{Experimental Setup:} Across all experiments, three distinct types of distributed nodes are defined and deployed within the emulated network environment as listed below. 
Communication between the distributed nodes is via point-to-point bidirectional links implemented with a DropTail queue. 


\begin{itemize}[leftmargin=7mm,topsep=2pt]
\item {\bf writer $w \in W \subseteq C$ :} 
a client that 
dispatches update requests to servers.

\item {\bf reader $r \in R \subseteq C$:} 
a client that 
dispatches read requests to servers 

\item {\bf server $s \in S$:} listens for reader and writer requests and is responsible for maintaining the object replicas according to the underlying 
protocol
they implement.
\end{itemize}

\noindent\textbf{Performance Metrics:} 
We assess performance using: (i) \emph{operational latency}, and (ii) \emph{the update success ratio}. The operational latency
is computed as the sum of communication and computation delays. In the case of \frfs{}, computational latency encompasses the time necessary for the {FM} to fragment a 
file object and generate the respective hashes for its blocks. The update success ratio is the percentage of update operations that have not been converted to reads (and thus successfully changed the value of the indented 
object). In the case of \vmwABD{}, we compute the percentage of successful updates on the file as a whole over the number of all updates. For \frfs{}, we compute the percentage of file updates, where all individual block updates succeed.

\noindent\textbf{Scenarios:} Both algorithms are evaluated under the following experimental scenarios:

\begin{itemize}[leftmargin=7mm,topsep=2pt]
  \item {\bf Scalability:} examine performance as the number of service participants increases 
  \item {\bf File Size:} examine performance when using 
  different initial file sizes 
  \item {\bf Block Size}: examine performance under 
  different block sizes (\frfs{} only)
\end{itemize}

We use a \emph{stochastic} invocation scheme in which reads are scheduled randomly from the intervals $[1...rInt]$ and updates from $[1..wInt]$, where $rInt, wInt = 4sec$. 
%
To perform a fair comparison and to yield valuable observations, the results shown are compiled as averages over five samples per each scenario.


\noindent\textbf{Scalability Experiments:}
We varied the number of readers $|R|$, the number of writers $|W|$, and the number of servers $|S|$ in the set $\{5, 10, 15, 20, 25, 30, 35, 40, 45, 50\}$. While testing for readers' scalability, the number of writers and servers was kept constant, $|W|,|S|=10$. Using the same approach, scalability of writers, and in turn of servers, was tested while preserving the two other types of nodes constant (i.e. $|R|,|S|=10$ and $|R|,|W|=10$ respectively). In total, each writer performed 20 updates and each reader 20 reads. The size of the initial file used was set to \SI{18}{\kilo\byte}, while the maximum, minimum and average block sizes (\emph{rabin fingerprints} parameters) were set to \SI{64}{\kilo\byte}, \SI{2}{\kilo\byte} and \SI{8}{\kilo\byte} respectively.
 
\noindent\textbf{File Size Experiments:}
We varied the $f_{size}$ from \SI{1}{\mega\byte} to \SI{1}{\giga\byte} by doubling the 
file size in each simulation run. 
The number of writers, readers and servers was fixed to 5. In total, each writer performed 5 updates and each reader 5 reads. The maximum, minimum and average block sizes 
were set to \SI{1}{\mega\byte}, \SI{512}{\kilo\byte} and \SI{512}{\kilo\byte} respectively.


\noindent\textbf{Block Size Experiments:} We varied the minimum and average $b_{sizes}$ of \frfs{} from \SI{1}{\kilo\byte} to \SI{64}{\kilo\byte}. 
The number of writers, readers and servers was fixed to 10. In total, each writer performed 20 updates and each reader 20 reads. The size of the initial file used was set to \SI{18}{\kilo\byte}, while the maximum block size was set to \SI{64}{\kilo\byte} 


%

\begin{figure}[!p]
{\small \centering
\begin{tabular}{cc}
	\includegraphics[scale=0.5,width=0.5\textwidth,height=50mm]{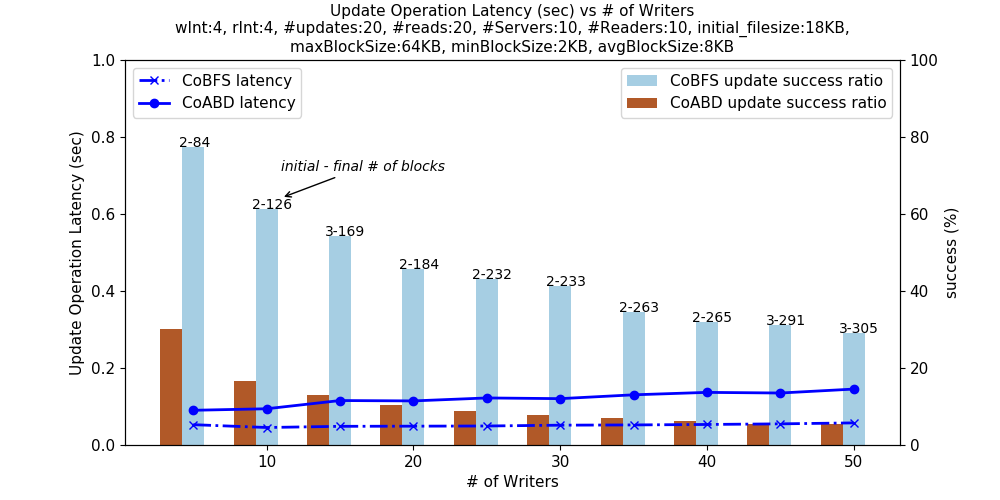}
	&
	\includegraphics[scale=0.5,width=0.5\textwidth,height=50mm]{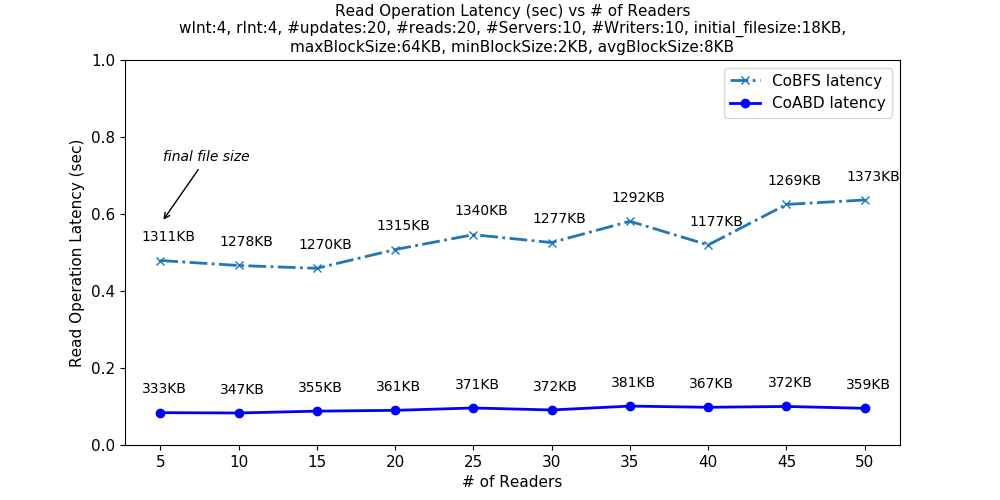}\\
	(a)  & (b) \\
	\includegraphics[scale=0.5,width=0.5\textwidth,height=50mm]{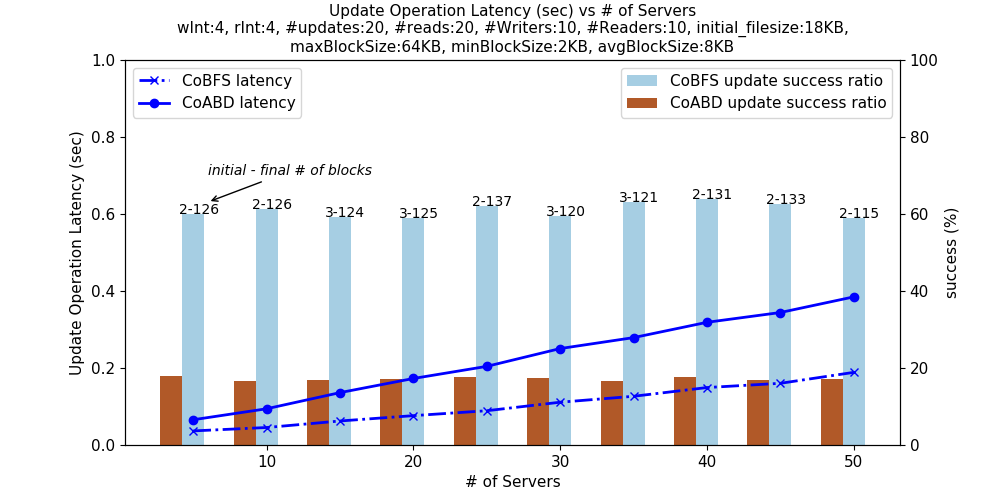}
	&		 
	\includegraphics[scale=0.5,width=0.5\textwidth,height=50mm]{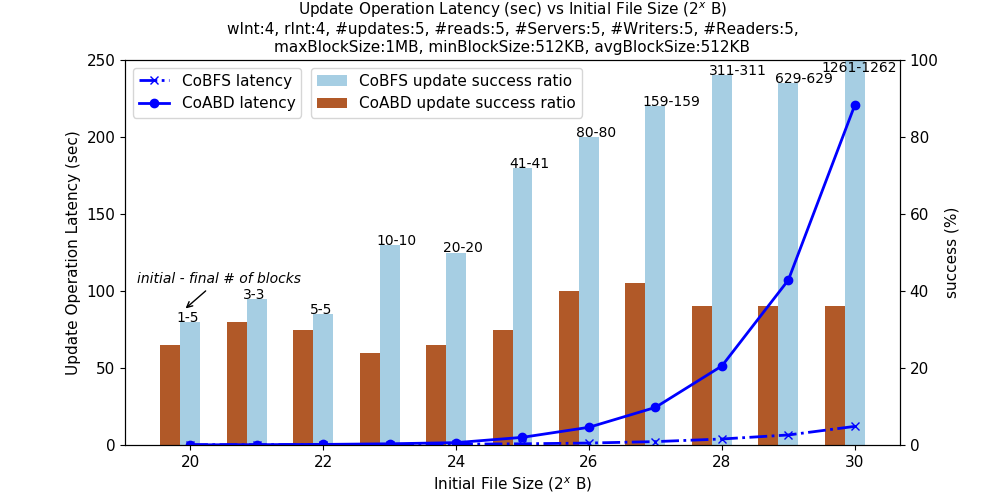} \\
	(c) & (d) \\ 
	\includegraphics[scale=0.5,width=0.5\textwidth,height=50mm]{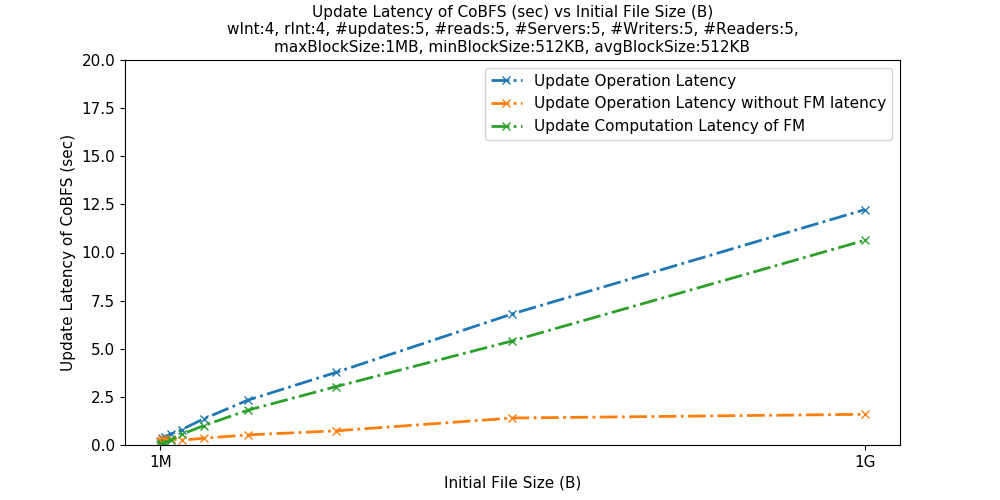} 
	&
	\includegraphics[scale=0.5,width=0.5\textwidth,height=50mm]{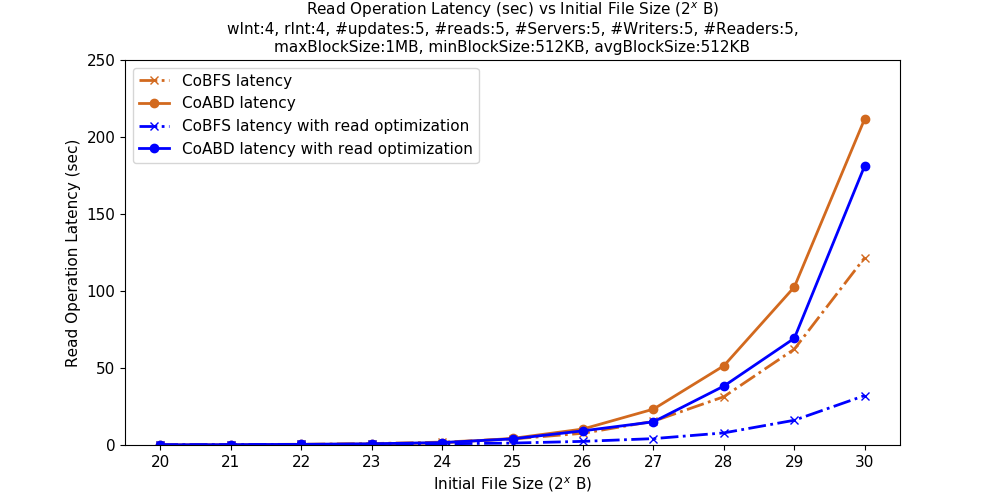} \\
	(e) & (f) \\ 
	\includegraphics[scale=0.5,width=0.5\textwidth,height=50mm]{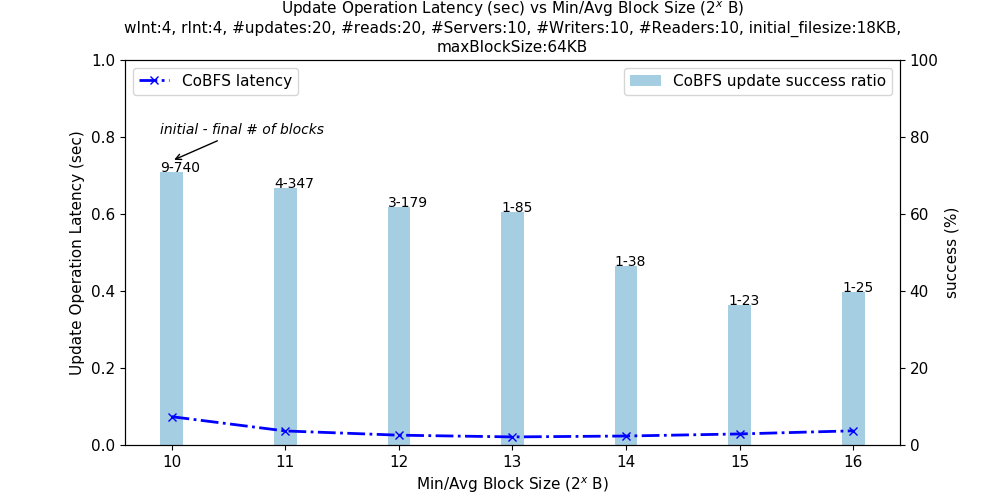}
	&
	\includegraphics[scale=0.5,width=0.5\textwidth,height=50mm]{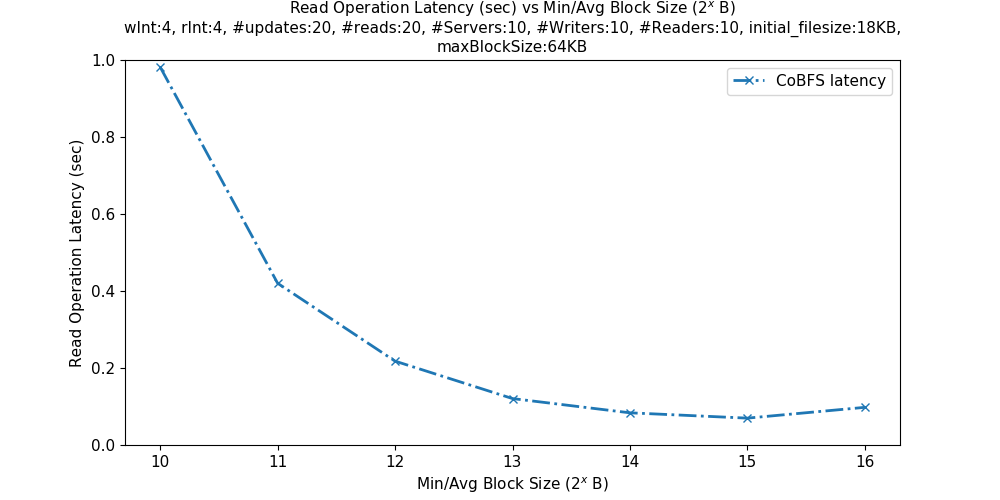} \\
	(g) & (h) \\ 
	
\end{tabular}
}
\vspace{-1em}
\caption{
Simulation results for algorithms \vmwABD{} and \frfs{}.
}
\label{fig:plots}
\end{figure}

\noindent\textbf{Results:} 
Overall, our results suggest that the efficiency of \frfs{} is inversely proportional to the number of block operations, rather than the size of the file. This is primarily due to the individual block-processing nature of \frfs{}. More in detail:
\vspace{.3em} 

\noindent\textit{Scalability:}
In Fig.~\ref{fig:plots}(a), the operational latency of updates in \frfs{} remains almost unchanged and smaller than of \vmwABD{}. This is because \vmwABD{} writer updates a rather small file, while each \frfs{} writer updates a subset of blocks which are modified or created. The computational latency of {FM} in \frfs{} is negligible, when compared to the total update operation latency, because of the small file size.
In Fig.~\ref{fig:plots}(c), we observe that the update operation latency in \vmwABD{} increases even more as the number of servers increases.
As more updates are successful in \frfs{}, reads may transfer more data compared to reads in \vmwABD{}, explaining their slower completion as seen in
Fig.~\ref{fig:plots}(b). 
Also, readers send multiple read block requests of small sizes, waiting each time for a reply, while \vmwABD{} readers wait for a message containing a small file. 
\vspace{.3em} 

\noindent\textit{Concurrency:}
The percentage of successful file updates achieved by \frfs{} are significantly higher than those of \vmwABD{}. This holds for
both cases where the number of writers increased (see Fig.~\ref{fig:plots}(a)) and the number of servers increased (see Fig.~\ref{fig:plots}(c)). This demonstrates the boost of concurrency achieved by \frfs{}. 
In Fig.~\ref{fig:plots}(a) we notice that as the number of writers increases (hence, concurrency increases), \vmwABD{} suffers greater number of unsuccessful updates, i.e., updates that have become reads per the coverability property. 
Concurrency is also affected when the number 
of blocks increases, Fig.~\ref{fig:plots}(d). The probability of two writes to collide on a single block decreases, and thus \frfs{} eventually allows all the updates (100\%) to succeed. 
\vmwABD{} does not experience any 
improvement as it always manipulates the file as a whole.\vspace{.3em}

\noindent\textit{File Size:}
%
Figure~\ref{fig:plots}(d) demonstrates that the update operation latency of \frfs{} remains at extremely low levels. The main factor that significantly contributes to the slight increase of \frfs{} update latency is the FM computation latency, 
Fig.~\ref{fig:plots}(e). 
We have set the same parameters for the \emph{rabin fingerprints} algorithm for all the initial file sizes, which may have favored some file sizes but burdened others. An  optimization of the rabin algorithm or a use of a different algorithm for managing blocks could possibly lead to improved FM computation latency; this is a subject for future work. The \frfs{} update communication latency remains almost stable, since it depends primarily on the number and size of update block operations. That is in contrast to the update latency exhibited in \vmwABD{} which appears to increase linearly with the file size. This was expected, since as the file size increases, it takes longer latency to update the whole file. 

Despite the higher success rate of \frfs{}, the read latency of the two algorithms is comparable due to the low number of update operations. The read latencies of the two algorithms with and without the read optimization can be seen in Fig.~\ref{fig:plots}(f).  The \vmwABD{} read latency increases sharply, even when using the optimized reads. This is 
in line with our initial hypothesis, as \vmwABD{} requires reads to request and propagate the whole file each time a newer version of the file is discovered. 
Similarly, when read optimization is not used in \frfs{}, the latency is close of \vmwABD{}. Notice that each read that discovers a new version of the file 
needs to request and propagate the content of each individual block. On the contrary, read optimization decreases significantly the \frfs{} read latency, as reads transmit only the contents of the blocks that have changed.\vspace{.3em}

\noindent\textit{Block Size:}
From Figs.~\ref{fig:plots}(g)(h) we can infer that when smaller blocks are used, the update and read latencies reach their highest values. In both cases, small $b_{size}$
results in the generation of larger number of blocks from the division of the initial file. Additionally, as seen in Fig.~\ref{fig:plots}(g), the small $b_{size}$ leads to the generation of more new blocks during
\act{update} operations, resulting in more update block operations, and hence higher latencies. 
As the minimum and average $b_{sizes}$ increase, lower number of blocks need to be added when an \act{update} is taking place. Unfortunately, smaller number of blocks leads  to a lower success rate. 
Similarly, in Fig.~\ref{fig:plots}(h), smaller block sizes require more read block operations to obtain the file's value.
As the minimum and average $b_{sizes}$ increase, lower number of blocks need to be read. Thus, further increase of the minimum and average $b_{sizes}$ forces the decrease of the latencies, reaching a plateau in both graphs. This means that the emulation finds optimal minimum and average $b_{sizes}$ and increasing them does not give better (or worse) latencies.

%% file: conclusions.tex
\section{Conclusions}
\label{sec:conclude}

We have introduced the notion of linearizable and coverable fragmented objects and 
proposed an algorithm 
that implements coverable fragmented files. 
It is then used to build
\frfs{}, a prototype distributed file system in which each file is specified as a linked-list of coverable blocks. \frfs{} adopts a modular architecture, separating the object fragmentation process from the shared memory service 
allowing
to follow different fragmentation strategies and shared memory implementations. We showed that it preserves the validity of the fragmented object (file) and satisfies fragmented coverability.
The deployment 
on Emulab serves as a proof of concept implementation.
The evaluation 
demonstrates the potential of our approach in boosting the concurrency and improving the efficiency of R/W operations on strongly consistent large objects. 

For future work, we aim 
to perform a comprehensive experimental evaluation of \frfs{}
that will go beyond simulations (e.g., full-scale, real-time, cloud-based experimental evaluations) 
and to further study parameters that may affect the performance of the operations (e.g., file size, block size, etc), as well as to build optimizations and extensions, in an effort to unlock the full potential of our approach. 



%% file: appendix_v1.tex
\newpage
\appendix
\section*{\LARGE Appendix}

\section{Fragmented Objects with Coverable Blocks}
\label{sec:coverable}

When writing a value to a linearizable R/W object, the value written does not need to be dependent on the previous written value. However, in some objects (e.g. files), it is
expected that a value update will build upon (and thus avoid to overwrite) the current value of the object. 
In such cases a writer should be aware of the latest value of the object (i.e., by reading the object) before updating it. 
Although a read-modify-write (RMW) semantic would be more appropriate for this type of objects, it can only be achieved through consensus, which is known to be merely impossible to solve in 
an asynchronous environment with crashes \cite{FLP85}.  


To this respect, in~\cite{NFG16} the notion of {\em coverability} was introduced
to leverage the solvability of R/W object implementations, while providing a \emph{weak} RMW object. Informally, coverability, extends linearizability with the additional guarantee that object writes succeed when associating the written value 
with the ``current'' {\em version} of the object. In a different case, a write operation becomes a read operation and returns the latest version and the associated value of the object.  

More formally, coverability uses a \textit{totally ordered} set of \textit{versions}, say $\verSet$, and introduces the notion of \emph{versioned (coverable) objects}. 
A \emph{coverable object} is a type of R/W object where each value written 
is assigned with a version from the set $\verSet$. 
The \emph{coverable} R/W object $X$ offers two operations:
 (i) $X.\act{cvr-write}(\val{}, ver)_{\pr}$, and (ii) $X.\act{cvr-read}()_{\pr}$. 
A process $\pr$ invokes a $\act{cvr-write}(\val{}, ver)_{\pr}$ operation 
when it performs a write operation that attempts to change the value 
of the object. 
The operation returns
the value of the object and its associated version, along with a flag informing
whether the operation has successfully changed the value of the object or failed.
A write is \emph{successful} if it changes the value of the 
register; otherwise the write is \emph{unsuccessful}.
The read operation $\act{cvr-read}()_{\pr}$ involves a request to 
retrieve the value of the object. The response of this operation is the 
value of the register together with the version of the object that this value is 
associated with. Denoting a successful write $\act{cvr-write}(v, ver)(v, ver', chg)_{\pr}$ as $\trw{ver}{ver'}_{\pr}$ (updating the object from version $ver$ to $ver'$),
and $\act{cvr-write}(v, ver)(v', ver', unchg)_{\pr}$ as $\trw{ver}{ver', unchg}_{\pr}$, a coverable implementation satisfies the following properties (for the formal definition see~\cite{NFG16}). 

\begin{definition}[Coverability~\cite{NFG16}]
\label{def:weak}
A valid execution $\EX$ is \textbf{coverable} with respect to a total order $<_{\EX}$ 
on all successful write operations,$\wSet_{\EX,succ}$, in $\EX$ if:\vspace{-.5em}
\begin{itemize}[leftmargin=7mm]\itemsep2pt
	\item ({\bf Consolidation}) If a $\trw{ver_j}{*}\in\wSet_{\EX,succ}$ then $ver_j$ is larger than any version written by a preceding successful write operation.
	\item ({\bf Continuity}) if $\trw{ver}{ver_i}\in\wSet_{\EX, succ}$, then $ver$ was written by a preceding write operation or $ver=\bot$ the initial version
	\item ({\bf Evolution}) The version of the object is incrementally evolving and thus for two version `chains' formed by concurrent writes on a single initial version $ver$, the last version of the longest chain is larger than the latest version on the shorter chain.
\end{itemize}
\end{definition}

If a fragmented object utilizes coverable blocks, instead of linearizable blocks, then Definition~\ref{def:fragatomic} provides what we would call {\bf\em fragmented coverability}: Concurrent update operations on different blocks would \emph{all} prevail (as long as each update is tagged with the latest version of each block), whereas only one update operation on the same block would prevail (all the other updates on the same block that are concurrent with this would become a read operation). As we 
see in the next section fragmented coverability is a good alternative to RMW semantics to implement large objects, like files, of which any new value may depend on the current value of the object.

\section{Additional Operations Supported by the Prototype}
To enhance the practicality of our prototype we have equipped it with   additional operations, which are all framed around the two main operations of the FM. 

Besides updating the contents of a file, reading a file and managing blocks, the {\em FM} supports a number of other useful operations, such as creating a file, renaming a file, deleting a file, obtaining a list of the existing files and an advanced list operation. 
\newline
%

To store information about the files that the \emph{FM} manages, internally the \emph{FM} maintains a dictionary $D$. In more detail, a key entry is a file path $f_{path}$ of $f_{id}$, and the corresponding value is a tuple consisting the $b_{id}$ of the genesis block $b_{g}$ of $f_{id}$ and  the file id $f_{id}$ of the fragmented file $f$. That is, $D: \{ key,value \} = \{ f_{path} , \langle b_{g}, f_{id}   \rangle \}$.

The \emph{FM} uses $f_{path}$ as key for this dictionary, in order to be able to monitor the changes that take place for each file. However, in the level of the Atomic Shared Object Algorithm, all the information about a file is stored based on its $f_{id}$. 

It is worth mentioning that, the format of a block that sending to the Atomic Shared Object Algorithm, is a dictionary containing the header and the literal data of the block. The header includes some information about the block, i.e. the hash value, a boolean value that indicates if the block is the genesis one, the next $b_{id}$, the block size and the modification time of the block. If the block is the genesis block, the header it also contains the $f_{path}$. 

\begin{itemize}[leftmargin=10mm]


\item {\bf Create Operation}: When a new file is created on the client's filesystem, 
the \emph{FM} 
fragments it into its respective blocks (including the genesis block), and writes them on the servers by invoking a sequence of write operations for the entirety of the blocks comprising the file. 

\item {\bf Rename Operation}: When a file is renamed on the client, the \emph{FM} executes a special write request, where it writes the genesis block of the file that includes the new $f_{path}$ in its header. 

\item {\bf Delete Operation}: When a file is deleted on the client, the \emph{FM} discards the $f_{id}$ entry from its dictionary and sends a special write request to the servers, with the genesis bid $b_{gen}$ of the file. The servers set the tag of the $b_{gen}$ to -1, in order to notify that the file is deleted in case another client tries to have access to it before the delete operation is completed. As a result, no further operations can be performed on the deleted file, since the \emph{FM} and the servers do not have access to its genesis block. 

\item {\bf List Operation}: To obtain the list of existing files, the {\em FM} contacts the servers and obtains the $f_{id}$, the $f_{path}$ and the genesis block id $b_{id}$ of each file, which then allows for further read operations to be issued.

\item {\bf Advanced List Operation}: The advanced list operation, is similar to the simple list one, giving some additional information about each file. At first, the \emph{FM} requests a simple list operation. Then for each file in the resulted list, it requests a series of block list operations. Each block list operation informs the \emph{FM} about the size and the modified size of the block. As a result, the \emph{FM} can calculate the size of the whole file and the maximum modified time that a block of the file has changed. 

\end{itemize}